\documentclass[]{aa} % for a referee version
\usepackage{graphicx}
\usepackage{txfonts}
\usepackage{xcolor}
\usepackage[normalem]{ulem}
\usepackage[amsmath,thmmarks]{ntheorem}
\theoremseparator{.}
\theorembodyfont{\normalfont} 
\newtheorem{assumption}{Assumption}
\DeclareMathOperator{\sinc}{sinc}
\begin{document} 

   \title{\textit{Kalkayotl}: A cluster distance inference code }
%   \subtitle{I. Distance determination}

   \author{J. Olivares\inst{1}
          \and L.~M. Sarro\inst{2}
          \and H. Bouy \inst{1}
          \and N. Miret-Roig\inst{1}
          \and L. Casamiquela\inst{1}
          \and P.A.B. Galli\inst{1}
          \and A. Berihuete\inst{3}
          \and Y. Tarricq\inst{1}
          }

   \institute{Laboratoire d'astrophysique de Bordeaux, Univ. Bordeaux, CNRS, B18N, allée Geoffroy Saint-Hilaire, 33615 Pessac, France.
         \email{javier.olivares-romero@u-bordeaux.fr}
         \and
         Depto. de Inteligencia Artificial , UNED, Juan del Rosal, 16, 28040 Madrid, Spain 
         \and
         Depto. de Estad\'i{}stica e Investigaci\'on Operativa , Universidad de C\'adiz, Campus Universitario R\'i{}o San Pedro s/n, 11510 Puerto Real, C\'adiz, Spain
         }

   \date{Received; accepted }

% \abstract{}{}{}{}{} 
% 5 {} token are mandatory
 
  \abstract
   {Stellar clusters are benchmarks for theories of star formation and evolution. The high precision
parallax data of the \textit{Gaia} mission allows significant improvements in the distance determination to stellar clusters and its stars. In order to have accurate and precise distance determinations, systematics like the parallax spatial correlations need to be accounted for, especially for stars in small sky regions.}  
   {Provide the astrophysical community with a free and open code designed to simultaneously infer cluster parameters (i.e. distance and size) and the distances to its stars using \textit{Gaia} parallax measurements. It includes cluster oriented prior families, and is specifically designed to deal with the \textit{Gaia} parallax spatial correlations.}
   {A Bayesian hierarchical model is created to allow the inference of both the cluster parameters and distances to its stars.}
   {Using synthetic data that mimics \textit{Gaia} parallax uncertainties and spatial correlations, we observe that our cluster oriented prior families result in distance estimates with smaller errors than those obtained with an exponentially decreasing space density prior. In addition, the treatment of the parallax spatial correlations minimizes errors in the estimated cluster size and stellar distances, and avoids the underestimation of uncertainties. Although neglecting the parallax spatial correlations has no impact on the accuracy of cluster distance determinations, it underestimates the uncertainties and may result in measurements that are incompatible with the true value (i.e. fall beyond the 2$\sigma$ uncertainties).}
   {The combination of prior knowledge with the treatment of \textit{Gaia} parallax spatial correlations produces accurate (error <10\%) and trustworthy estimates (i.e. true values contained within the 2$\sigma$ uncertainties) of clusters  distances for clusters up to $\sim$5 kpc, and cluster sizes for clusters up to $\sim$1 kpc.}

   \keywords{Methods:statistical,Virtual Observatory tools, Parallaxes, Galaxy:open clusters and associations, Star:distances}

   \maketitle
%
%-------------------------------------------------------------------

\section{Introduction}
Stellar clusters offer a unique opportunity to test models of the formation and evolution of stars and stellar systems. Their distance is useful to compare model predictions to observations when the observational uncertainties are taken into account. Traditionally, these comparisons use the cluster distance, which is typically more precise than that of individual stars. However, high-precision astrometric surveys, like \textit{Hipparcos} \citep{ESA1997,1997A&A...323L..49P} and \textit{Gaia} \citep{gaia,2018A&A...616A...1G}, have pushed these comparisons at the level of individual stars, at least for the most precise measurements of the nearest systems. Thus, while determining cluster distances remains a fundamental problem, retrieving the distances to the cluster stars allows astronomers to perform detailed tests of current theories of star formation and evolution; see for example the analyses of internal dynamics and 3D structures done by \citet{2018MNRAS.476..381W}, \citet{2019A&A...630A.137G}, and \citet{2020MNRAS.494.4794A}. 

Since the cluster distance is an important parameter, diverse methodologies have been developed to estimate it either from photometry, astrometry, or combinations of them \citep[e.g.][]{2014A&A...564A..49P,2015A&A...576A...6P,2017A&A...601A..19G,2017A&A...598A..48G,2018A&A...615A..12Y}. In the context of distance determination based on parallax measurements the traditional approach consists of averaging the parallaxes of the cluster stars and then inverting the resulting more precise parallax mean. More sophisticated approaches have also been devised. For example, \citet{2014A&A...564A..49P} developed a maximum-likelihood approach for open cluster distance determination. The authors assumed that the spatial distribution of stars in open clusters follows a spherical Gaussian distribution, and inferred the cluster distance together with its dispersion and other kinematic parameters by marginalizing the positions of individual stars. They validated their methodology on synthetic clusters with properties similar to those expected for the \textit{Gaia} data. \citet{2017A&A...601A..19G} determined astrometric parameters of open clusters by modeling their intrinsic kinematics and projecting them in the observational space. \citet{2018A&A...615A..49C} obtained open cluster distances using a maximum likelihood method. Nonetheless, they neglect the cluster intrinsic depth, which results in underestimated distance uncertainties.

Although devised outside the context of open clusters, the following Bayesian frameworks are worth of mention due to their use of \textit{Gaia} parallax measurements and specific prior information. \citet{2018AJ....156...58B} inferred posterior distance distributions to 1.3 billion stars in the \textit{Gaia} data using a Galactic weak distance prior specifically designed for the entire Galaxy. \citet{2019A&A...628A..94A} obtained distances to stars brighter than G=18 mag using a Galactic multi-component prior (with halo, bulge, thin and thick disks), and a combination of \textit{Gaia} measurements (including parallax) and photometry from several surveys. \citet{2018MNRAS.476..381W} developed a forward-model for the inference of parameters of OB associations. Their model uses a 3D \citet{1987ApJ...323...54E} profile that includes the distance to the association as a free parameter. However, they inferred the individual stellar distances using the Galactic distance prior proposed by \citet{2018AJ....156...58B}. In addition, the online resources\footnote{\url{https://github.com/agabrown/astrometry-inference-tutorials}, and \url{https://github.com/ehalley/parallax-tutorial-2018}.} recommended by \citet{2018A&A...616A...9L} provide detailed steps for the inference of cluster distance and size. Recently, \citet{2020A&A...637A..95P} used the insight provided by these online resources to infer cluster distances based on a Gaussian prior. These authors marginalize the individual distances as well as the cluster intrinsic dispersion.

The previous studies can be classified into those that infer population parameters of either clusters and associations by marginalizing individual stellar distances \citep[e.g.][and the online resources mentioned above]{2018MNRAS.476..381W,2020A&A...637A..95P}, and those that infer individual stellar distances but do not infer the population parameters \citep[e.g][]{2018AJ....156...58B,2019A&A...628A..94A}. To the best of our knowledge, the simultaneous inference of population parameters and individual distances has not been addressed in the literature. Furthermore, none of the aforementioned methodologies is able to deal with the systematics introduced by the parallax spatial correlations present in the \textit{Gaia} data \citep[see Sect. 5.4 of][]{2018A&A...616A...2L}.

Following the guidelines provided by \citet{2018A&A...616A...9L}, in this work we attempt to solve the aforementioned issues in the specific context of stellar clusters by providing the astrophysical community with the free open code \textit{Kalkayotl}\footnote{Kalkayotl means distance in the mesoamerican Nahuatl language.}. It samples the joint posterior distribution of the cluster parameters and stellar distances, given their \textit{Gaia} astrometric data and a set of cluster oriented prior families. 

Our approach is different from that adopted in the aforementioned works. While \citet{2018AJ....156...58B}, \citet{2018MNRAS.476..381W} and  \citet{2019A&A...628A..94A}  analytically or numerically find statistics of the stars posterior distance distributions using a unique Galactic prior, \textit{Kalkayotl} obtains samples of the posterior distribution of the star and cluster parameters for a set of cluster oriented prior families by means of Hamiltonian Monte Carlo \citep{1987PhLB..195..216D}, which is a type of Markov Chain Monte Carlo (hereafter MCMC) technique. This approach offers the user the advantage of taking an active criticism over the prior (i.e.
choose its family, infer its parameters, and compare with 
results from other prior families), together with an easier 
propagation of uncertainty into subsequent analyses. On the other hand, the MCMC approach has the constraint of being computationally expensive. In a machine with four CPUs at 2.7 GHz, \textit{Kalkayotl} takes typically five minutes to run the inference model of a cluster with one thousand stars, although the running time can increase depending on the prior complexity and quality of the data set.

The rest of this work is organized as follows. In Sect. \ref{section:methodology} we introduce the methodology of \textit{Kalkayotl}. In Sect. \ref{section:data_sets} we construct synthetic clusters that mimic the \textit{Gaia} data, and in Sect.\ref{section:validation} we use these clusters to validate the methodology. Finally, in Sect. \ref{section:conclusions} we discuss the advantages and caveats of our methodological approach and present our conclusions.

\section{Methodology}
\label{section:methodology}

\textit{Kalkayotl} is a free python code\footnote{The code and documentation is available at: \url{https://github.com/olivares-j/Kalkayotl}} designed to simultaneously sample the joint posterior distribution of cluster parameters and stellar distances. In addition, the user can decide to only sample the stellar distances by fixing the cluster parameters, and/or perform the sampling on the parallax space (i.e. sampling the cluster and sources true parallaxes). The latter can be useful when the subsequent analyses need to be done on the parallax space. Although the methodology can be applied to any parallax measurement, \textit{Kalkayotl} is specifically designed to work with \textit{Gaia} astrometric data. The users will also be able to run the code through the Spanish Virtual Observatory, although with limitations.

\subsection{Assumptions}
\label{subsection:assumptions}
Previous to describing the details of the methodology, we state the undertaken assumptions.

\begin{assumption}
\label{ass:gaussian_uncertainties}
The \textit{Gaia} astrometric measurements are normally distributed around the true values. As explained in Sect. 5.2 of \citet{2018A&A...616A...2L}, the standardized astrometric measurements are almost\footnote{The standard deviations of the standardized parallax, and proper motions in R.A. and Dec., are 1.081, 1.093 and 1.115, respectively. Thus the errors are 8-12\% larger than the formal uncertainties.} normal. The \textit{Gaia} catalog provides all necessary information (i.e. mean, standard deviations, and correlations) to reconstruct these distributions.
\end{assumption}

\begin{assumption}
\label{ass:zero_points}
The \textit{Gaia} parallax measurements are shifted from their true value, and this shift can be different for different {sky positions}, colors, and magnitudes \citep{2018A&A...616A...1G}. Extensive studies have been made to determine this parallax zero point and its correlations with other observables and {stellar types} \citep[see Fig. 14 of][and references therein]{2019arXiv191000398C}. The user can set the parallax zero point value\footnote{The parallax zero point uncertainty can be included by adding it to the parallax uncertainty of the sources.}, and if desired use different values for different sources.
\end{assumption}

\begin{assumption}
\label{ass:correlations}
The \textit{Gaia} astrometric measurements of different sources are spatially correlated amongst them. We use the covariance functions proposed by \citet{2019MNRAS.tmp.2030V}, which provide a better {description to the observed correlations at small angular separations than those of \citet{2018A&A...616A...2L}; compare Fig. 2 of the former author to Fig. 15 of the latter authors. The parallax covariance function of \citet{2019MNRAS.tmp.2030V}} is given by:
\begin{equation}
    V(\theta)=0.0003\cdot\exp(-\theta/20\degr) + 0.002\cdot\sinc(0.25\degr + \theta/0.5\degr)\ \ \rm{mas}^2
\end{equation}
where $\theta$ is the angular separation between two sources.
\end{assumption}

\begin{assumption}
\label{ass:proportion}
The cluster size is much smaller than its distance. If the cluster size is comparable to its distance then using only the line of sight distances results in biased estimates of the cluster distance due to projection effects.
\end{assumption}

\begin{assumption}
\label{ass:selection_function}
The input list of cluster members is neither contaminated nor biased (i.e. we assume a perfect selection function).
\end{assumption}

As in any Bayesian methodology, we now proceed to specify the likelihood, prior and the procedure to obtain the posterior distribution.

\subsection{Likelihood}
\label{sect:likelihood}
The likelihood of the $N$ observed sources with data $\boldsymbol{\mathcal{D}}=\{\varpi_i,\sigma_{\varpi,i}\}_{i=1}^{N}$ (where $\varpi_i$ and $\sigma_{\varpi,i}$ are the mean and standard deviation that define the parallax measurement of source $i$), given parameters $\boldsymbol{\Theta}$, can be represented as

\begin{equation}
    \mathcal{L}(\boldsymbol{\mathcal{D}}|\boldsymbol{\Theta})\equiv \mathcal{L}(\{\varpi_i\}_{i=1}^N|T(\boldsymbol{\Theta}),\{\sigma_{\varpi,i}\}_{i=1}^{N})=\mathcal{N}(\mathtt{X}-\mathtt{X}_{zp}|T(\boldsymbol{\Theta}),\mathtt{\Sigma}), \label{equation:multivariate_likelihood}
\end{equation}

where $\mathcal{N}(\cdot|\cdot)$ represents the multivariate normal distribution (see Assumption \ref{ass:gaussian_uncertainties}), $\mathtt{X}$ the $N$-dimensional vector of the observed parallax, $\mathtt{\Sigma}$ the $N \times N$ covariance matrix, $\mathtt{X}_{zp}$ the $N$ vector of zero points, and $T$ the transformation from the parameter space to the {space of} observed-quantities. 

In the set of parameters $\boldsymbol{\Theta}=\{\theta_i\}_{i=1}^N$, the parameter $\theta_i$ of the $i^{\rm{th}}$ source represents its true distance. Therefore, $T$ is equal to $1000/\theta$, with $\theta$ in pc  and the result in mas. If the user decides to do so, the sampling can be done in the parallax space, in which case $\theta_i$ corresponds to the source true parallax, and $T$ is the identity relation. Nonetheless, hereafter we work in the distance space.

The vectors $\mathtt{X}$ and $\mathtt{X}_{zp}$ are constructed from the concatenation of the $N$ vectors of observations $\{\varpi_i\}_{i=1}^N$ and zero-points $\varpi_{zp}$ respectively (see Assumption \ref{ass:zero_points}).   

The covariance matrix $\mathtt{\Sigma}$ contains the $N$-dimensional vector of variances, $\{\sigma_{\varpi,i}^2\}_{i=1}^{N}$, in its diagonal, whereas the off-diagonal terms are the covariances between the parallax measurements of different sources (see Assumption \ref{ass:correlations}).

\subsection{Prior families}
\label{subsection:priors}
The prior distribution is supposed to encode the previous knowledge of the investigator about the plausibility of the parameter values in a model. In the case of stellar clusters, we know that their stars share {common distributions of their} astrophysical properties, like their distance to the observer, age, metallicity, etc.. This \textit{a priori} information is what we use to construct an informed prior. Nonetheless, given the variety of cluster morphologies, we believe that there is no universal prior for clusters.

In \textit{Kalkayotl} we propose two types of distance prior families, one based on classical statistical probability density distributions and another based on purely astrophysical considerations. The purely statistical ones are common distributions used in the literature, while the astrophysical ones are inspired by previous works devoted to the analysis of the luminosity (or number) surface density profiles of galactic and globular clusters. The purely statistical prior families are parametrized only by their location, $loc$, and scale, $scl$, which are defined as follows. The location, $loc$, is the expected value of the cluster distance, while the scale, $scl$, is the typical scale length of the cluster along the line of sight. 
Our statistical prior families are the following. The Uniform prior family is the simplest one, as it assigns the same probability density to all values in the interval $[loc-scl,loc+scl]$. The Gaussian prior family assumes that the distance is normally distributed with mean, $loc$, and standard deviation, $scl$. The Gaussian Mixture Model prior family (hereafter GMM) assumes that the distance distribution is described by a linear combination of $k$ Gaussian distributions, with $k$ an integer greater than zero. In the following, we will use $k=2$. We also analyzed other types of distributions (like Cauchy, Half-Cauchy, and Half-Gaussian) but they returned poorer results when compared to the previously mentioned prior families, and thus we do not include them in our analysis.

The astrophysical prior families are parametrized as well by the location $loc$ and scale $scl$ parameters, but they contain more than these, as will be described below. The $loc$ parameter still describes the most typical cluster distance, while the $scl$ one now corresponds to what is commonly referred to as the core radius (i.e. the typical size of the cluster inner region). It is important to notice that although the astrophysical distance prior families have similar functional forms to the luminosity (or number) surface density profiles from which they were inspired, there is no correspondence between them; while the latter are defined as surface densities, the former are defined as distance densities. 

The Elson, Fall, and Freeman prior family (hereafter EFF) distributes the distances in a similar way as \citet{1987ApJ...323...54E} distributed the surface luminosity density of clusters from the Large Magellanic Cloud. In addition to the location and scale parameters, it utilizes the $\gamma$ parameter which describes the slope of the distribution at large radii. In the standardized form ($loc=0$ and $scl=1$), the EFF is defined as:

\begin{equation}
    {\rm EFF}(r|\gamma)=\frac{\Gamma(\gamma)}{\sqrt{\pi}\cdot \Gamma(\gamma-\frac{1}{2})}\cdot \left[ 1 + r^2\right]^{-\gamma},
\end{equation}
with $\Gamma$ the gamma function, and $r$ the standardized distance. In our parametrization $\gamma=\gamma'/2$, with $\gamma'$ the original parameter proposed by \citet{1987ApJ...323...54E}. We notice that by fixing the $\gamma$ parameter to 1 or 5/2 the EFF prior family reduces to the Cauchy and Plummer distributions, respectively.

The King prior family distributes the distances in a similar way as \citet{1962AJ.....67..471K} distributed the surface number density of globular clusters. In addition to the location and scale parameters, it includes the maximal extension of the cluster through the tidal radius parameter, $r_t$; the probability distribution is thus normalized within this distance. In its standardized form ($loc=0$, core radius=$scl=1$), the King prior is defined as
\begin{equation}
    {\rm King}(r|r_t)=\frac{
	   \left[
	   \frac{1}{\sqrt{1 + r^2}}-\frac{1}{\sqrt{1 + r_t^2}}
	   \right]^2}{2\left[\frac{r_t}{1+r_t^2} -\frac{2{\rm arcsinh}(r_t)}{\sqrt{1+r_t^2}}+\arctan(r_t)\right]},
\end{equation}
with $r$ the standardized distance. 

For completeness reasons, \textit{Kalkayotl} also includes the Galactic exponentially decreasing space density prior (hereafter EDSD) prior\footnote{We do not refer to it as a prior family since its only parameter, the scale length, will be kept fixed throughout this analysis.} introduced by \citet{2015PASP..127..994B}. We refer the interested reader to the aforementioned work for an explicit definition of this prior. Suffices to say that its only parameter, $scl$, is the typical length of exponential decay. In this work our objective is to estimate distances to stars in clusters, and not in the field population, thus we include it only for comparison purposes. We are perfectly aware that this will constitute an unfair comparison but we want to emphasize the problems associated with adopting a Galactic prior for the inference of distances in a cluster scenario.

\subsection{Hyper-priors}
\label{subsection:hyper_priors}
If the user decides to infer the cluster parameters, $\phi=\{loc,scl\}$, together with the source parameters $\boldsymbol{\Theta}$, then a hierarchical model is created with the cluster parameters at the top of the hierarchy. In this case, a prior must be set for each parameter of the chosen cluster prior family. In the Bayesian jargon, this kind of prior is called hyper-prior, and its parameters, hyper-parameters.

As hyper-prior for the location, $loc$, and scale, $scl$, we use the Normal($loc$|$\boldsymbol{\alpha}$) and Gamma($scl$|2,2/$\beta$) densities  respectively, where $\boldsymbol{\alpha}$ and $\beta$ are their hyper-parameters. The Gamma distribution and its hyper-parameters are specified following the recommendations of \citealp[][]{Chung2013}. The specific choice of the rate parameter as 2/$\beta$ in the Gamma distribution results in the mean of the latter at $\beta$.

The weights, $\{w_i\}_{i=1}^k$ in the GMM prior family are Dirichlet( $\{w_i\}_{i=1}^k | \boldsymbol{\delta})$ distributed, with $\boldsymbol{\delta}$ the $k$-th vector of hyper-parameters. The $\gamma$ parameter in the EFF prior family is distributed as $\gamma \sim 1+{\rm Gamma}(2,2/\gamma_{hyp})$ with $\gamma_{hyp}$ provided by the user; this parametrization avoids $\gamma$<1, which will produce extreme cluster tails. For the tidal radius in the King prior family we use a similar weakly informative prior: $r_t \sim 1+{\rm Gamma}(2,2/\gamma_{hyp})$ with $\gamma_{hyp}$ an hyper-parameter provided by the user. We notice that the tidal radius is in units of the core radius (i.e. scale parameter), and thus it is restricted to be larger than one.

\subsection{Posterior distribution}
\label{subsection:posterior}

Bayes' theorem states that the posterior distribution equals the prior times the likelihood normalized by the evidence $\boldsymbol{\mathcal{Z}}$. If the user of  \textit{Kalkayotl} decides to infer only the distances to the individual sources, this is only the $\boldsymbol{\Theta}$ parameters, then the posterior distribution is given by
\begin{equation}
\label{equation:posterior_0}
    \mathcal{P}(\boldsymbol{\Theta} \mid \boldsymbol{\mathcal{D}})=\frac{\mathcal{L}(\boldsymbol{\mathcal{D}}\mid \boldsymbol{\Theta})\cdot \pi(\boldsymbol{\Theta}\mid \phi)}{\boldsymbol{\mathcal{Z}}},
\end{equation}
where the likelihood $\mathcal{L}$ is given by Eq.\ref{equation:multivariate_likelihood}. The prior $\pi$ is one of the prior families described in Sect. \ref{subsection:priors} for which its parameters, $\phi$ have been fixed to a user decided value. Finally, the evidence $\boldsymbol{\mathcal{Z}}$ is simply the normalization factor. 

On the other hand, if the user decides to infer both the source distances, $\boldsymbol{\Theta}$, and the cluster parameters, $\phi$, then the posterior is given by
\begin{equation}
\label{equation:posterior_1}
    \mathcal{P}(\boldsymbol{\Theta},\phi \mid \boldsymbol{\mathcal{D}})=\frac{\mathcal{L}(\boldsymbol{\mathcal{D}}\mid \boldsymbol{\Theta})\cdot \pi(\boldsymbol{\Theta}\mid \phi)\cdot \psi(\phi)}{\boldsymbol{\mathcal{Z}}},
\end{equation}
where now $\psi$ is the hyper-prior of the cluster parameters $\phi$. We notice that this latter case should not be used in combination with the EDSD prior because it will be meaningless to infer the scale length of a Galactic prior based on data from the population of a single Galactic cluster.

In \textit{Kalkayotl} the posterior distribution is sampled using the Hamiltonian Monte Carlo method implemented in \textit{PyMC3} \citep{2016ascl.soft10016S}, which is a Python probabilistic programming framework. For details of the capabilities and caveats of Hamiltonian Monte Carlo samplers on hierarchical models we refer the reader to the work of \citet{2013arXiv1312.0906B}. In particular, our model faces the typical problems of sampling efficiency associated with hierarchical models. Thus, following the recommendations of the aforementioned authors, \textit{Kalkayotl} allows the user to choose between the central and non-central parametrizations (except for the GMM prior family, which contains more than one scale parameter). While the non-central parametrization enables more efficient sampling in the presence of poorly-informative data sets (i.e. few members and/or large uncertainties), the central one works better for more constraining ones (i.e. nearby and well-populated clusters).

\textit{PyMC3} provides different initialization schemes for the MCMC chains, and a set of tools to automatically diagnose convergence after sampling. We choose the \texttt{advi+adapt\_diag} initialization scheme\footnote{\label{footnote:pymc3}The interested reader can find more details about initialization schemes and convergence diagnostics at the \textit{PyMC3} documentation: \url{https://docs.pymc.io/}}, because it proved to be the most efficient one to reduce both the number of tuning steps (thus the total computing time) and initialization errors (like those of "Bad initial energy" or "zero derivative" for a certain parameter). This method starts the chain at the test value (which depends on the prior but is usually its mean or mode) and runs the automatic differentiation variational inference algorithm, which delivers an approximation to the target posterior distribution.

Once the initialization is completed, the code performs the inference in two stages. First, the sampler is tuned, and then the posterior samples are computed. The number of tuning and sampling steps are chosen by the user. While sampling steps are established based on the desired parameters precision\footnote{The parameter precision is given by the standard error of the mean: $\sigma/\sqrt{n}$, where $\sigma$ is the posterior standard deviation, and $n$ its effective sample size (i.e. number of independent samples from the posterior distribution). This last value is reported by the sampler and is proportional to the input value of sampling steps given by the user and the sampler efficiency. For details of its computation see \url{https://mc-stan.org/docs/2_18/reference-manual/effective-sample-size-section.html} }, the number of tuning steps depends on the complexity of the posterior. Typical values of the tuning steps are 1000 and 10000 for the simple (i.e. Uniform and Gaussian) and complex (i.e. EFF, King, and GMM) prior families, respectively.

Once the inference is finished, the convergence of the chains is assessed based on the Gelman-Rubin statistic, effective sample size, and the number of divergences (see Note \ref{footnote:pymc3}). \textit{Kalkayotl} then discards the tuning samples (to avoid biased estimates) and reports cluster and source summary statistics (the desired percentiles and the mode, median or mean). In addition, it also makes trace plots of the cluster and source parameters. Although the automatic analysis made by \textit{PyMC3} usually suffices to ensure convergence, we strongly recommend the users to visually inspect the chains to ensure that no anomalies are present.

As in most Bayesian inference problems, the investigator must face the decision of choosing the most suitable prior. The rule of thumb is that there is no universal prior, and the most suitable prior depends on the specific problem at hand. Thus, to help users decide which prior family might be the most suitable for their data sets, \textit{Kalkayotl} offers a module to make comparison of models by means of Bayes factors, which are the ratio of the Bayesian evidence ($\boldsymbol{\mathcal{Z}}$ in Eqs. \ref{equation:posterior_0} and \ref{equation:posterior_1}) of each pair of models. Estimating the Bayesian evidence is a hard and computationally expensive problem, thus, in order to reduce the computation time extra assumptions are needed. The reader can find these extra assumptions together with details of the evidence computation in Appendix \ref{appendix:evidence}.  Once the Bayesian evidence of each model is computed, the decision can be taken based on \citet{Jeffreys61} scale\footnote{\label{footnote:Jeffreys}In this scale, the evidence is: inconclusive if the Bayes Factor
is <3:1, weak if it is $\sim$3:1, moderate if it is $\sim$12:1, and strong if it is > 150 :1.}.

Summarizing, \textit{Kalkayotl} returns samples of the joint posterior distribution of the cluster parameter and stellar distances, together with summary statistics thereof. In addition, the user can do model selection based on the Bayes factors. However, we notice that the evidence computation is expensive, taking at least three and ten times more time than the posterior sampling of the purely statistical and astrophysical prior families, respectively.

\section{Synthetic clusters}
\label{section:data_sets}
Our aim in this section is to create synthetic clusters with parallax uncertainties and spatial correlations similar to those present in the \textit{Gaia} data. These synthetic clusters will then be used to validate our methodology, in particular its accuracy and precision as a function of cluster distance and the number of sources.

The \textit{Gaia} parallax uncertainty depends on the source magnitude, colour and number of transits \citep[see][]{2018A&A...616A...1G,2018A&A...616A...2L}. Thus, to generate realistic parallax uncertainties we simulate the photometry of our sources with the \textit{isochrones} python package \citep{2015ascl.soft03010M}. The mass of each source was randomly drawn from a Chabrier mass distribution \citep{2005ASSL..327...41C} and its photometry computed by means of the MIST models \citep{2016ApJS..222....8D,2016ApJ...823..102C}. For the latter, we use solar metallicity, zero extinction, and the typical open cluster age of 158 Myr, which corresponds to the mean age of the 269 open clusters analyzed by \citet{2019A&A...623A.108B}. We explore a grid of distances from 100 pc to 1 kpc at steps of 100 pc, and from 1 kpc to 5 kpc at steps of 1 kpc. We use 100, 500, and 1000
sources, these cover the typical numbers of clusters members. 

The radial distance of each source to the cluster center was drawn from each of our distance prior families, and its 3D Cartesian coordinates were then computed. We notice that these result in spherically symmetric distributions, which suffices for the purposes of the present analysis.

To account for random fluctuations, we repeat ten times each simulation of our grid (main distance, number of sources, and distance distribution). We use ten parsecs as the typical cluster scale, and for the EFF and King prior we set their $\gamma$ and standardized tidal radius parameters to 3 and 5, respectively. In the GMM synthetic clusters, we use for the second component a distance 10\% larger than the main distance, and a scale of 20 pc. The fraction of sources in each component was set to 0.5.

Then, we use \textit{PyGaia}\footnote{\url{https://github.com/agabrown/PyGaia}} to obtain parallax uncertainties (from the G, and V-I photometry together with nominal GDR2 time baseline) and to transform the true source coordinates into true sky positions and parallaxes. 

Afterward, we use sky positions, parallax uncertainties, and \citet{2019MNRAS.tmp.2030V} parallax spatial correlation function to compute the covariance matrix $\Sigma$, which is constructed by adding the covariance matrix of the parallax uncertainties (i.e. a diagonal matrix with parallax variances in the diagonal) plus the covariance matrix of the parallax spatial correlations (see Assumption \ref{ass:correlations}). Then, the observed parallaxes were drawn from a multivariate normal distribution centered on the true parallaxes and with $\Sigma$ as the covariance matrix. We did not include any parallax zero-point shift in our synthetic data sets. We end up with a total of 2100 synthetic clusters containing 1.12 million sources.

\section{Validation}
\label{section:validation}

In this section, we measure the accuracy, precision, and credibility of our
methodology at estimating the true values of both the population and source parameters (further details and additional figures can be found in Appendix \ref{appendix:quality}). We measure accuracy and precision as the fractional error (i.e. the posterior mean minus the true value divided by the true value) and the fractional uncertainty (i.e. the 95\% posterior credible interval divided by the true value), respectively. We define credibility as the percentage of synthetic clusters realizations in which the inferred 95\% posterior credible interval contains the true value (i.e. the true value is covered by the 2$\sigma$ uncertainties). This definition of credibility measures the trustworthiness of the inferred value and its reported uncertainty. In this section, we also compare the distance estimates delivered by different prior families when applied to the same synthetic cluster. Furthermore, we analyze the sensitivity of our methodology to the choice of hyper-parameter values, the detail of this analysis can be found in Appendix \ref{appendix:sensitivity}. Briefly, we find that the results of our methodology are insensitive to changes of up to 10\% and 50\% in the hyper-parameters of the location and scale parameters, respectively.

Additionally, we use our set of synthetic clusters to explore the accuracy, precision, and credibility of the commonly used approach of inverting the mean parallax of the cluster stars. The results of this analysis are shown in Appendix \ref{appendix:naive}. Briefly, we observe that this approach returns cluster distance estimates with low fractional errors (<5\%) when the cluster is located closer than 1 kpc. However, beyond that limit, the approach is susceptible to large random errors (>10\%), as already reported by \citet{2014A&A...564A..49P}. Moreover, the low uncertainties obtained by inverting the mean parallax, result in smaller credibilities than those obtained by our methodology over the same data sets (compare Fig. \ref{fig:naive} with the left column of Fig. \ref{fig:A&P_Gaussian}). The only exception being the closest clusters, at 100 pc, where the validity of our Assumption \ref{ass:proportion} is the weakest.

\subsection{Accuracy and precision}
\label{subsection:accuracy_and_precision}

Concerning the population parameters, we find that the cluster distance is accurately determined by all our prior families, with a fractional error smaller than 10\%. The cluster scale accuracy depends on the chosen prior family, the number of cluster sources, and the cluster distance. This parameter is accurately determined, with a fractional error smaller than 10\%, by the Uniform, Gaussian, and King prior families in clusters located up to 0.7-1 kpc. However, the EFF and GMM prior families show fractional errors that are systematically larger than 20\%. Furthermore, the GMM prior family showed convergence problems in cluster beyond 1 kpc. 

The performance of the prior family at recovering the true parameter values is directly related to its complexity (the number of parameters is a good proxy for it). The Uniform and Gaussian prior families produce the lowest fractional errors, the smallest uncertainties, and the largest credibility. The King family also attains large credibility in its parameters despite its low identifiability\footnote{A model is said to be identifiable when different parameter values generate different observed distributions (i.e the model is non-degenerate).}. The latter is caused by the tidal radius, in which different and large values of it produce similar distance distributions in the central region particularly. The EFF prior family has a degeneracy between its scale and $\gamma$ parameters resulting in low credibility and large fractional errors. Finally, the GMM produces the lowest credibility among all prior families and the largest fractional errors in the location parameter. For all these reasons, we encourage the users of \textit{Kalkayotl} to perform inferences in order of prior complexity: start with the Uniform and Gaussian families and move to the King, EFF, and GMM ones only if needed.

Our results show that neglecting the parallax spatial correlations has negative consequences. Although neglecting these correlations has no major impact on the accuracy of the location parameter (at least for clusters located closer than 4 kpc), it results in underestimated uncertainties; an effect already reported by \citet{2019MNRAS.tmp.2030V}. As a consequence, out of the ten realizations of each synthetic cluster, neglecting the parallax spatial correlations reduces the parameter credibility from more than 80\% to less than 60\% on average. Furthermore, neglecting these correlations results in systematically large fractional errors in the scale parameter of cluster located beyond 300 pc. In summary, neglecting the parallax spatial correlations lowers the credibility of both location and scale parameters.

The results about the accuracy, precision, and credibility of our methodology at recovering the individual source distances are summarized as follows. The accuracy is better than 3\% for all cluster distances and number of sources. The precision is better than 5\% in clusters closer than 1 kpc, and grows up to 15\% for the farthest ones, up to 5 kpc. The high precision and low uncertainty result in the high credibility, > 90\%, of our distance estimates. Neglecting the parallax spatial correlations increases the fractional errors, and thus diminishes the credibility of the estimates.

\subsection{Prior comparison}
\label{subsection:prior_comparison}

We finish our analysis by comparing the results obtained with different prior families on the same synthetic cluster. For simplicity, we show only the results of the synthetic cluster containing 500 stars, generated using the Gaussian distribution, and located at 500 pc. In addition, and for the sake of completeness, we also obtain distances with the EDSD prior. For the latter we use: i) a scale parameter of 1.35 kpc \citep{2016ApJ...832..137A}, and ii) following \citet{2018A&A...616A...9L} recommendations, we summarize the distance estimates of this prior using the mode of the posterior distribution. We run our methodology both including and neglecting the parallax spatial correlations. In both cases, the parallax zero point was set to 0 mas since our generated synthetic clusters do not include this offset. 

We find that all our cluster oriented prior families return trustworthy (i.e. true value contained within the 2$\sigma$ uncertainties) measurements of the cluster distance with fractional errors smaller than 1\%. The only exception is the GMM prior family, in which the fractional error of the cluster distance is 4\%. The main difference in the performance of the cluster prior families is at the source distance level, which is discussed below.

\begin{figure}[!ht]
    \centering
    \includegraphics[width=\columnwidth,page=1]{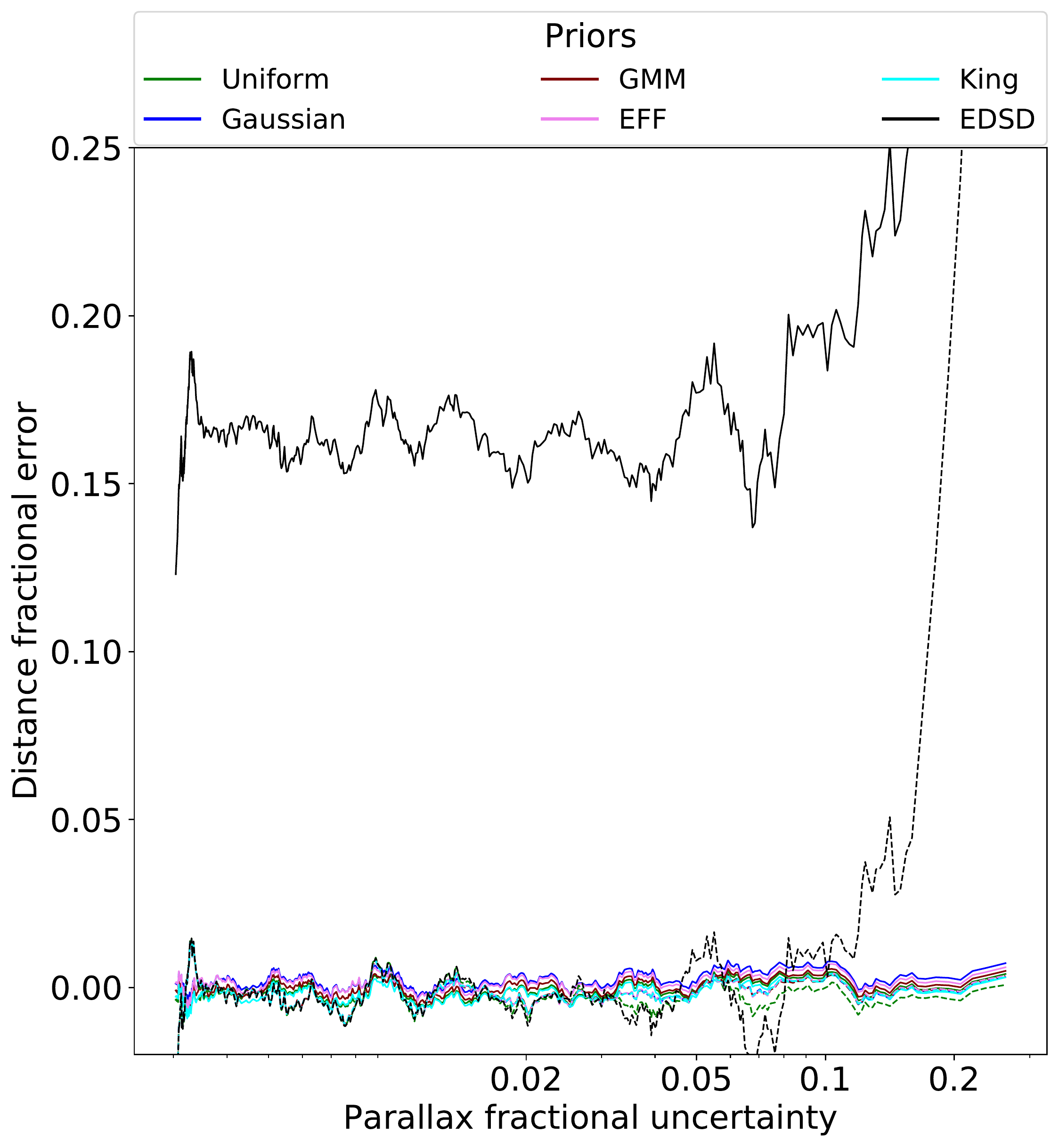}
    \caption{Distance fractional error as a function of parallax fractional uncertainty. The inference was done using all our prior families (color coded) on a synthetic cluster with 500 stars located at 500 pc. The lines show the rolling mean (computed with a window of 20 sources) of results obtained including (solid line) and neglecting (dashed lines) the parallax spatial correlations. }
    \label{fig:comparison_of_priors_0}
\end{figure}

Figure \ref{fig:comparison_of_priors_0} shows the rolling mean (with a window of 20 sources) of the fractional error in the inferred source distances as a function of their parallax fractional uncertainty. Distances were obtained using all our prior families plus the EDSD one, in each case the parallax spatial correlations were included and neglected (shown in the figure as solid and dashed lines, respectively). It is clear from this figure that neglecting the parallax spatial correlations when using the EDSD prior results in smaller fractional errors than those obtained when the parallax spatial correlations are taken into account, even for high-quality sources. The smaller fractional error results from neglecting the parallax spatial correlations, which is equivalent to assume that the data set is more informative than it is. In general, the more informative the data set is, the less influence the prior has on the posterior. Thus, when parallax spatial correlations are taken into account the mode of the posterior is attracted to the mode of the prior, which is located at 2.7 kpc (corresponding to 2$L$, with $L$ its length-scale, \citealp{2018AJ....156...58B}), hence the larger fractional error. Since the mode of the cluster oriented prior families is inferred from the data, it results in smaller fractional errors. 

\begin{table}[!ht]
    \caption{Fractional errors in the source distances. The columns show the prior family and the rms of the fractional error for three bins of the parallax fractional uncertainty. The number in parenthesis correspond to values obtained when the parallax spatial correlations are neglected. The last column shows the logarithm of the Bayesian evidence computed for each model.}
    \label{tab:bias}
    \resizebox{\columnwidth}{!}{
    \begin{tabular}{ccccc}
\hline
\hline
Prior  &  $f_\varpi$<0.05  &  0.05<$f_\varpi$<0.1  &  $f_\varpi$>0.1  &  log Z  \\
\hline
Uniform  &  6.98(12.36)  &  8.58(10.67)  &  10.82(11.16)  &  $103.15\pm0.16$  \\
Gaussian  &  6.88(11.29)  &  8.49(8.76)  &  10.87(10.88)  &  $103.57\pm0.17$  \\
GMM  &  6.90(11.29)  &  8.53(8.66)  &  10.89(10.90)  &  $103.07\pm0.18$  \\
EFF  &  6.98(11.26)  &  8.46(8.87)  &  10.77(10.81)  &  $103.92\pm0.17$  \\
King  &  6.89(11.28)  &  8.47(8.72)  &  10.85(10.90)  &  $104.52\pm0.15$  \\
EDSD  &  90.29(14.08)  &  94.73(34.40)  &  425.59(384.89)  &    \\
\hline
\end{tabular}
    }
\end{table}

Table \ref{tab:bias} shows the rms fractional error of the inferred distances for three different ranges of parallax fractional uncertainties. In the most precise parallax bin, that of $f_\varpi<0.05$, the Gaussian prior returns the smallest fractional error, followed closely by the King, GMM, EFF, and Uniform prior families. This result was expected since the true underlying distribution was Gaussian. In the less precise parallax bins ($f_\varpi>0.05$), the lowest fractional errors are those obtained with the EFF and King prior families. This interesting result shows that our astrophysical prior families produce excellent estimates of the source distances even when they do not match the true underlying distribution. In Table \ref{tab:bias}, the numbers in parentheses correspond to the rms of the fractional errors obtained when the parallax spatial correlations are neglected. These values are consistently larger than those obtained when the parallax spatial correlations are included. The only exception being the EDSD prior, for which the decrease in information produces a shift in its mode, as explained above. The lower fractional errors obtained by the cluster oriented prior families when the parallax spatial correlations are taken into account are the result of the proper modeling of the data characteristics. Finally, the last column of Table \ref{tab:bias} shows the logarithm of the Bayesian evidence computed for each of our cluster oriented prior families (the Bayesian evidence of the EDSD prior cannot be computed since its only parameter remains fixed). These Bayesian evidences are all very similar, and according to the Jeffreys scale (see Note \ref{footnote:Jeffreys}), their resulting Bayes factors provide inconclusive evidence to select one model over the others. Given the previous results, we can safely say that all our cluster oriented prior families are performing similarly well at recovering the source distances.

\begin{figure}[!ht]
    \centering
    \includegraphics[width=\columnwidth,page=3]{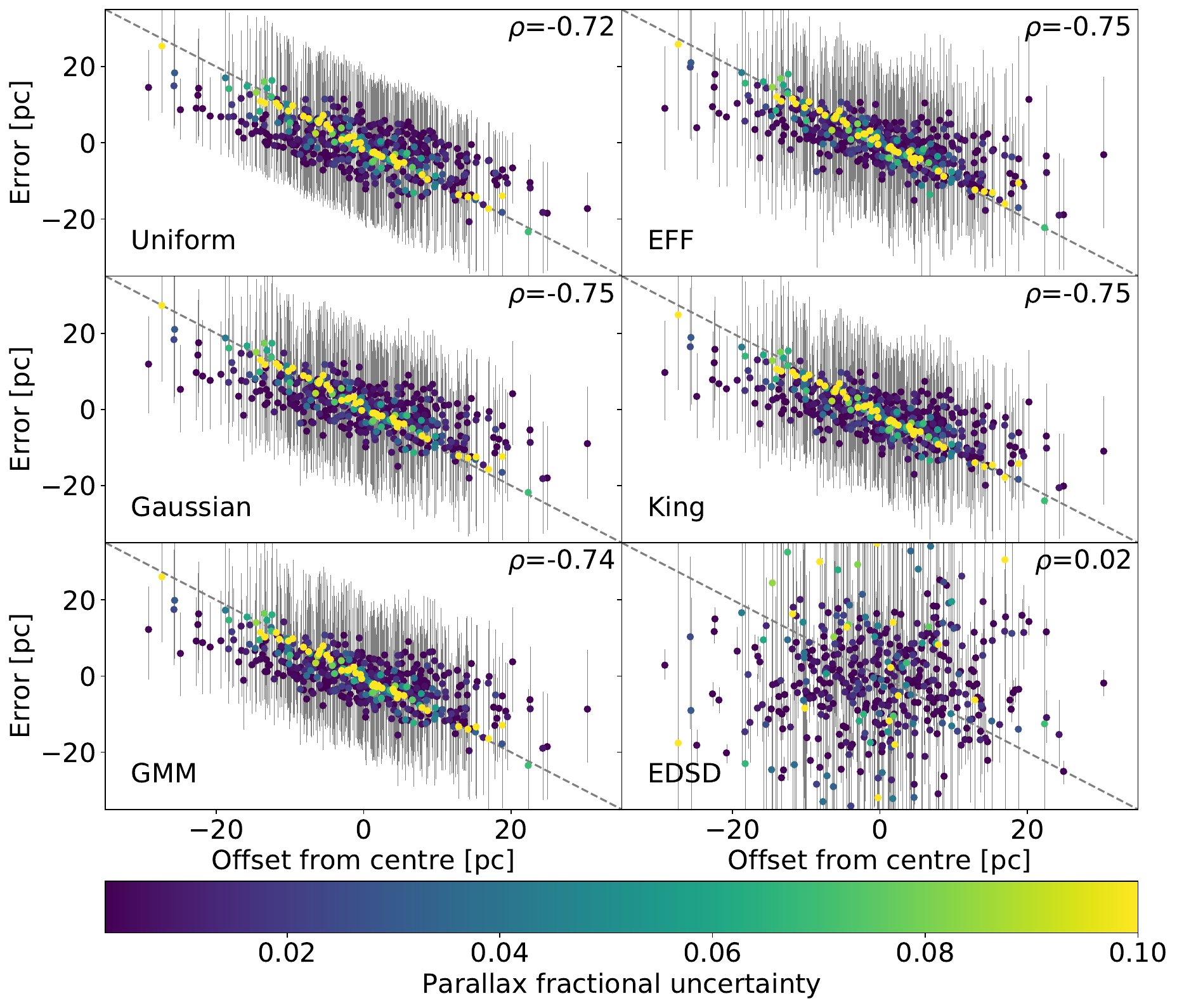}
    \caption{The error of the estimated source distances as a function of the offset from the cluster center (i.e. true distance minus cluster location). The panels show the results of different prior families. The color scale indicates the parallax fractional uncertainty and the gray dashed lines show the perfect anti-correlation. The Pearson correlation coefficient is shown on the top right corner of each panel.}
    \label{fig:comparison_of_priors_2}
\end{figure}

Despite the good performance of the cluster oriented prior families, they produce random errors in the source distance estimates that are inherent to the quality of the data. In Figure \ref{fig:comparison_of_priors_2} we show, for each prior family, the resulting distance error of individual sources as a function of its true position within the cluster. The distances obtained with all the cluster oriented prior families include the parallax spatial correlations, however, to make a fair comparison in the case of the EDSD prior they do not take into account these correlations. As can be observed, the distances obtained with the cluster oriented prior families show a clear anti-correlation between their error and the source offset to the cluster center (the Pearson correlation coefficients is shown in the top right corner of each panel of the figure). In the EDSD prior the correlation is negligible. The anti-correlation in the cluster prior families is proportional to the source fractional uncertainty; sources with large fractional uncertainty tend to fall over the dashed line of slope -1. 

The anti-correlation of this error has its origin in the same effect that produces the shift in the posterior distances obtained with the EDSD prior (see Fig. \ref{fig:comparison_of_priors_0}). In other words, when the parallax uncertainty is small, the information on the source location is more constraining and the prior plays a minor role. On the other hand, when the parallax uncertainty increases, its information reduces and the prior becomes more important. In this latter case, the posterior is attracted towards the mode of the prior, which results in the anti-correlation. 

From the comparison of the different prior families we conclude that: i) the cluster oriented prior families show an error that is proportional to the source fractional uncertainty, ii) the value of this error is smaller than that obtained with the EDSD prior (see Table \ref{tab:bias}). Thus we conclude that the cluster oriented prior families outperform the EDSD prior when inferring distances to stellar clusters. This comes as a no surprise since the EDSD prior was designed for the entire Galaxy and not for individual clusters. As explicitly mentioned by \citet{2015PASP..127..994B} "The exponentially decreasing volume density prior may be suitable when looking well out of the disk, where for a sufficiently deep survey the decrease in stellar density is caused mostly by the Galaxy itself rather than the survey." 

\section{Conclusions and future perspectives.}
\label{section:conclusions}

We make public the free and open code \textit{Kalkayotl}. It is a statistical tool for the simultaneous inference of star cluster parameters and individual distances of its stars. This tool utilizes distance prior families specifically designed for stellar clusters and takes into account the parallax spatial correlations present in the \textit{Gaia} data. Upon convergence, \textit{Kalkayotl} delivers high credibility (>90\%) estimates of distances to stellar clusters located closer than $\sim$5 kpc, and cluster sizes up to $\sim$1 kpc, with these values depending on the number of cluster stars. The samples from the posterior distributions of both cluster parameters and source distances can be used to propagate their uncertainties into subsequent analyses. 

Although the general formalism of our methodology can be applied to parallax measurements of diverse origins, our methodology is tuned to deal with the parallax spatial correlations of the \textit{Gaia} data. It is flexible enough to accommodate different values of parallax zero point and spatial correlation functions.

We validate this tool on realistic synthetic data sets and obtain the following conclusions:

\begin{itemize}
\item Distance estimates to sources with large fractional uncertainties (>0.05) can have large (> 20\%) systematic errors under incorrect assumptions. Provided that our assumptions are valid, these low-information sources can still be useful to constrain the cluster population parameters.

\item Compared to the inverse mean parallax approach, which results in cluster distance estimates that have low credibility (<80\%) but small fractional errors (< 5\%) for clusters up to $\sim$1 kpc, and high credibility (>80\%) but large fractional errors (>10\%) beyond this limit, our methodology returns distance estimates with small fractional errors (< 10\%) and high credibility (>90\%) for clusters up to $\sim$5 kpc. The exceptions are the nearest clusters ($\leq 100$ pc) and the GMM prior family.

\item The stellar distance estimates provided by \textit{Kalkayotl} show errors that are anti-correlated with the true position of the source relative to the cluster center. The anti-correlation is proportional to the source fractional uncertainty and reaches its maximum (-1) at distances larger than 1 kpc. Nonetheless, this error is still smaller than that incurred by the EDSD prior.

\item The spatial correlations in the parallax measurements are a non-trivial characteristic of \textit{Gaia} data. Neglecting them has negative consequences at both source and population level, amongst which, increased fractional errors, underestimated uncertainties, and low credibility are to be expected. Our results show that there is no objective reason in terms of accuracy, precision, or computing time to neglect the parallax spatial correlations when inferring the distances to clusters and its stars.

\item The amount of information provided by the data set is not always enough to constraint complex models. Thus we strongly suggest that the users of \textit{Kalkayotl} start with the simplest prior families (i.e. Uniform and Gaussian), verify their convergence, and later on, if needed, move to the more complex ones. If the latter are needed, their performance/convergence can be improved by reparametrizing: fixing some of the parameters or increasing the information content of the hyper-parameters. In this sense, users are encouraged to encode, by means of prior families and their hyper-parameters, the information they possess on the specific cluster that they analyze.

\end{itemize}

Although our methodology represents what we consider is an important improvement in the estimation of distances to stellar clusters from parallax data, it still has several caveats. Amongst those that we have detected and plan to address in the near future, we cite the following. 

\begin{itemize}
\item It is assumed that the list of cluster candidate members is not contaminated either biased (Assumption \ref{ass:selection_function}). However, in practice, this rarely happens. Cluster membership methodologies have certain true positive and contamination rates (see for example \citealp{2019A&A...625A.115O}). A further improvement of our methodology will be to simultaneously infer the cluster parameters and the degree of contamination while incorporating the selection function. 

\item The posterior distribution of the cluster distance may be further constrained by the inclusion of additional observations (e.g. photometry, proper motions, radial velocities, and sky positions). In the future, we plan to include the rest of the \textit{Gaia} astrometric observables to further constrain the parameters of stellar clusters.
\end{itemize}

\begin{acknowledgements}
We thank the anonymous referee for the comments that helped to improve the quality of this manuscript. This research has received funding from the European Research Council (ERC) under the European Union’s Horizon 2020 research and innovation program (grant agreement No 682903, P.I. H. Bouy), and from the French State in the framework of the ”Investments for the future” Program, IdEx Bordeaux, reference ANR-10-IDEX-03-02.
L.C. and Y.T. acknowledge support from "programme national de physique stellaire" (PNPS) and from the "programme national cosmologie et galaxies" (PNCG) of CNRS/INSU.
The figures presented here were created using Matplotlib \citep{Hunter:2007}.
Computer time for this study was provided by the computing facilities MCIA (Mésocentre de Calcul Intensif Aquitain)
of the Université de Bordeaux and of the Université de Pau et des Pays de l'Adour.
J.O expresses his sincere gratitude to all personnel of the DPAC for providing the exquisite quality of the \textit{Gaia} data.
\end{acknowledgements}

\bibliographystyle{aa} 
\bibliography{mybiblio.bib}

\begin{appendix}
\section{Bayesian evidence}
\label{appendix:evidence}
Here we provide details of the \textit{Kalkayotl} subroutine that computes the Bayesian evidence of the different prior families. Since \textit{PyMC3} does not provide evidence computation we use the python package \textit{dynesty} \citep{2019arXiv190402180S}.

The only purpose of this additional tool is to help the users of \textit{Kalkayotl} to decide which prior family is the most suitable to describe their data. Since here we are not interested in the distances to the stars, we marginalize them. The following provides the steps and assumptions undertaken in the marginalization of parameters $\boldsymbol{\Theta}$ from the posterior distribution (see Eq. \ref{equation:posterior_1}). Thus, the marginalization implies that
\begin{align}
    \mathcal{P}(\phi \mid \boldsymbol{\mathcal{D}}) &\equiv  \int \mathcal{P}(\boldsymbol{\Theta},\phi \mid \boldsymbol{\mathcal{D}})\rm{d}\boldsymbol{\Theta}\\ \nonumber
    &\propto \int \mathcal{L}(\boldsymbol{\mathcal{D}}\mid \boldsymbol{\Theta})\cdot \pi(\boldsymbol{\Theta}\mid \phi)\cdot \psi(\phi)\rm{d}\boldsymbol{\Theta}.
\end{align}
where the proportionality constant $\mathcal{Z}$ is the evidence that will be computed. To numerically compute the marginalization integral we made the following assumptions\footnote{We notice that these assumptions are only made for the sole purpose of evidence computation and they do not apply for the rest of \textit{Kalkayotl} methodology.}.

Assumption: the distances $\boldsymbol{\Theta} = \{\theta_i\}_{i=1}^N$ are independent and identically distributed with probability $\pi(\theta \mid \phi)$.

Assumption: the observed parallaxes $\boldsymbol{\mathcal{D}}=\{\varpi_i,\sigma_{\varpi,i}\}_{i=1}^{N}$ are independent (i.e. here we assume no spatial correlations) and normally distributed $\mathcal{N}(\cdot\mid\cdot)$. 

Under the previous assumptions, we have 

\begin{align}
    \mathcal{P}(\phi \mid \boldsymbol{\mathcal{D}})
    &=\psi(\phi)\cdot \prod_{i=1}^N \int \mathcal{N}(\varpi_i \mid T(\theta_i),\sigma_{\varpi,i})\cdot \pi(\theta_i \mid \phi)\cdot \rm{d}\theta_i.
\end{align}

Finally, we approximate each of the $N$ integrals by summing over a $M$-element $\{\theta_j\}_{j=1}^M$ of samples from the prior $\pi(\phi)$. Thus, we have
\begin{align}
    \mathcal{P}(\phi \mid \boldsymbol{\mathcal{D}})
    &\approx \psi(\phi)\cdot \prod_{i=1}^N \frac{1}{M}\sum_{j=1}^M \mathcal{N}(\varpi_i \mid T(\theta_j),\sigma_{\varpi,i}).
\end{align}

We run \textit{dynesty} with the following configuration. We use the \textit{static} Nested sampler with a single bound and a stopping criterion of $\Delta \log \mathcal{Z} < 1.0$ \citep[see][for more details]{2019arXiv190402180S}. To reduce the computing time we take a random sample of only $N=100$ cluster stars, always ensuring that this sample remains the same when computing evidence of different models. The $M$ value was set heuristically to 1000 prior samples.

\section{Details of the accuracy and precision}
\label{appendix:quality}

\begin{figure}[!ht]
    \centering
    \includegraphics[width=\columnwidth,page=1]{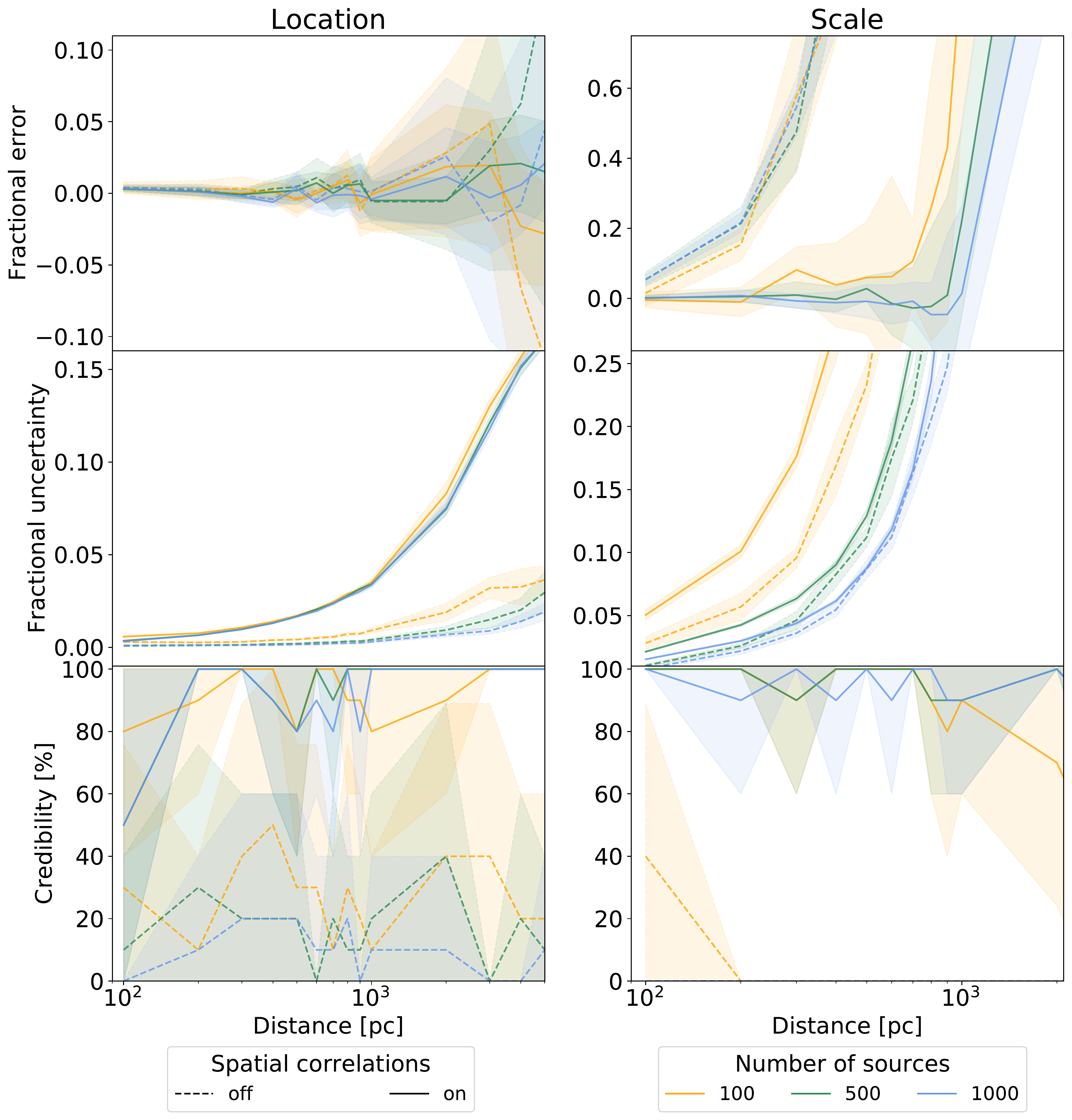}
    \caption{Fractional error, fractional uncertainty, and credibility of the population parameters as a function of distance. The parameters were inferred on the Uniform synthetic data sets using its corresponding prior family (i.e. Uniform), the colors indicate the number of sources, and the line styles show the cases in which the spatial correlations were included (solid) or were neglected (dashed). The lines show the mean of the ten simulations and the shaded areas its standard deviation.}
    \label{fig:A&P_uniform}
\end{figure}

\begin{figure}[!ht]
    \centering
    \includegraphics[width=\columnwidth,page=1]{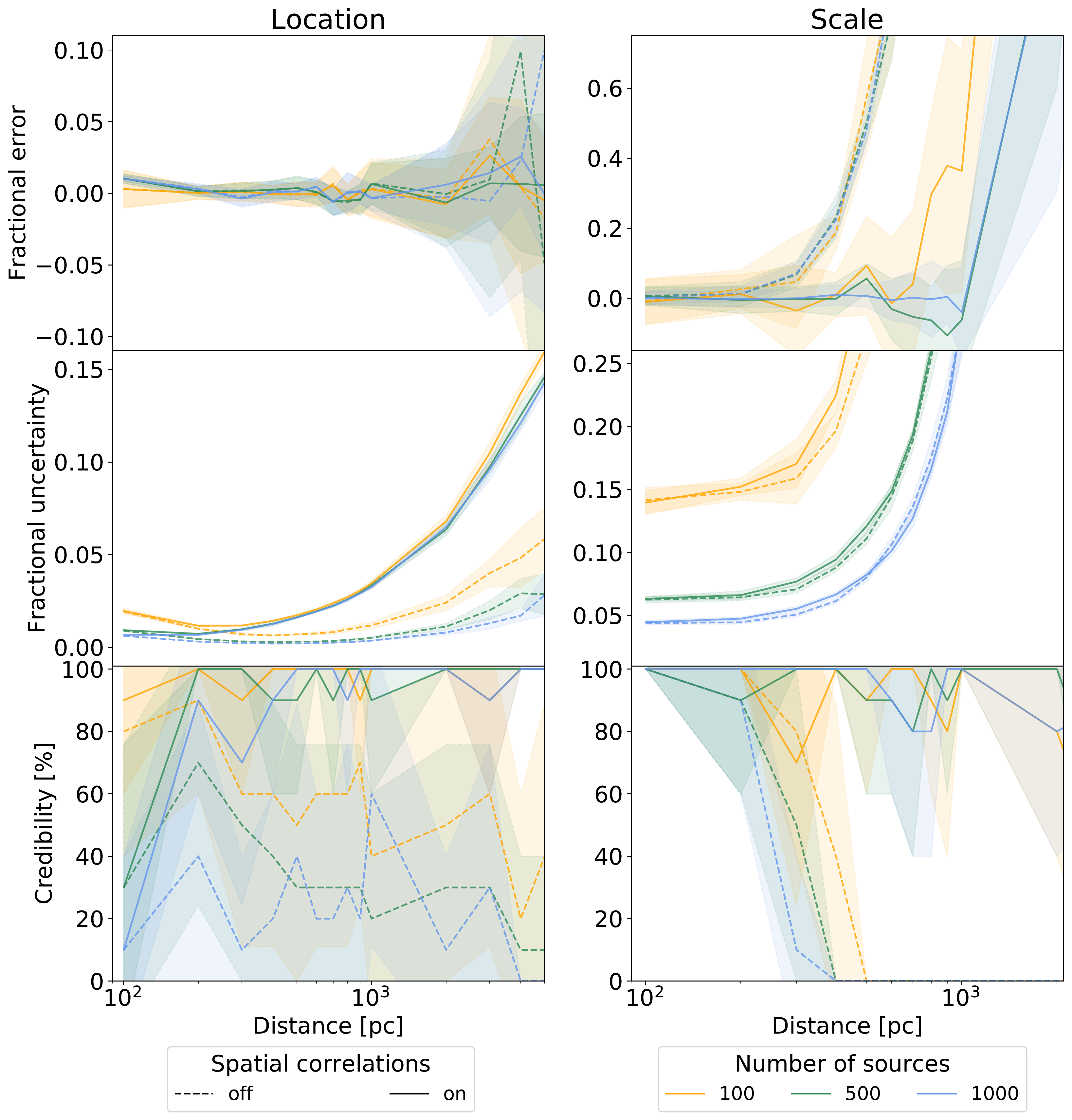}
    \caption{Same as Fig. \ref{fig:A&P_uniform} for the Gaussian prior.}
    \label{fig:A&P_Gaussian}
\end{figure}

\begin{figure}[!ht]
    \centering
    \includegraphics[width=\columnwidth,page=1]{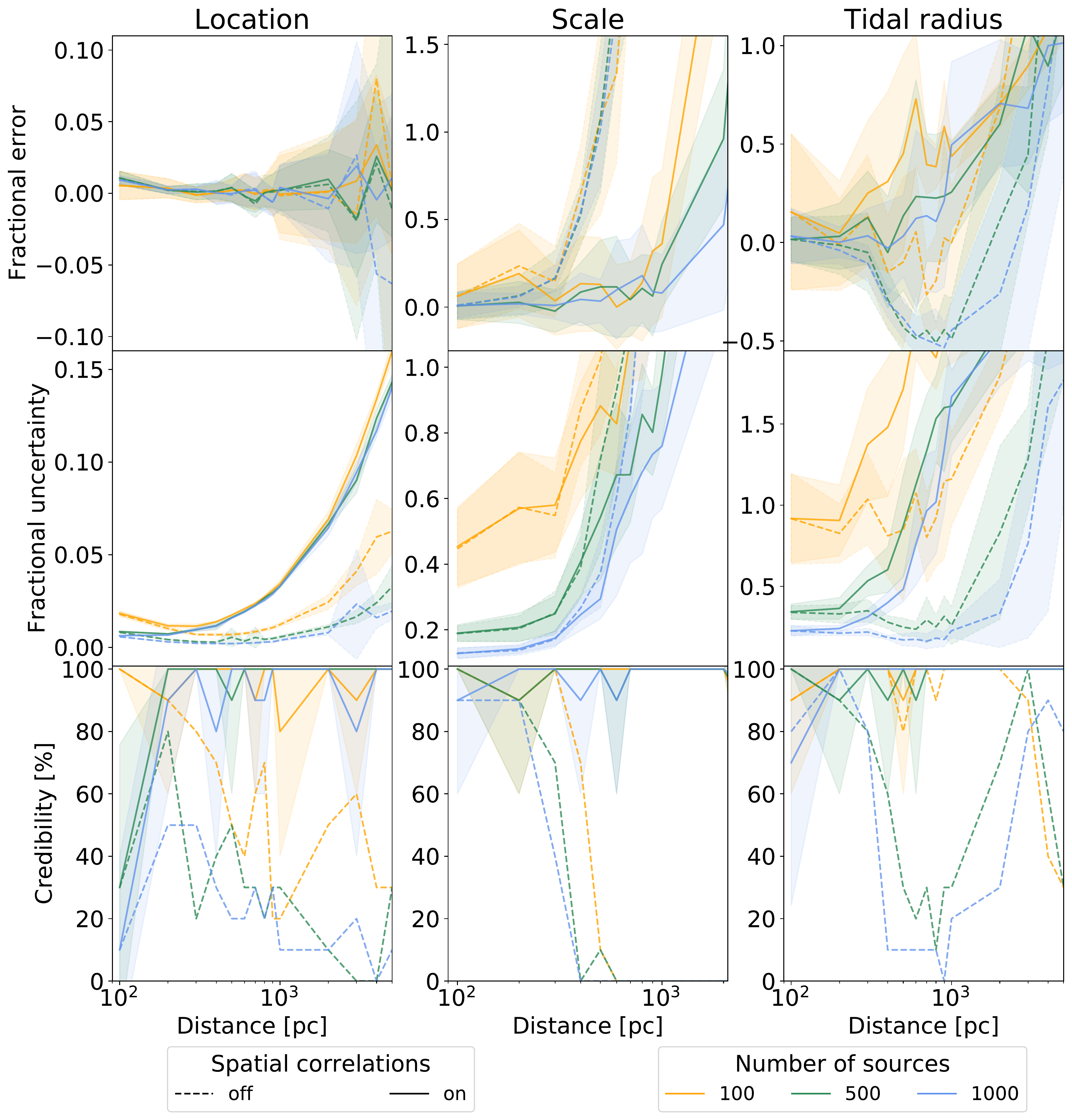}
    \caption{Same as Fig. \ref{fig:A&P_uniform} but for the King prior family.}
    \label{fig:A&P_King}
\end{figure}

\begin{figure}[!ht]
    \centering
    \includegraphics[width=\columnwidth,page=1]{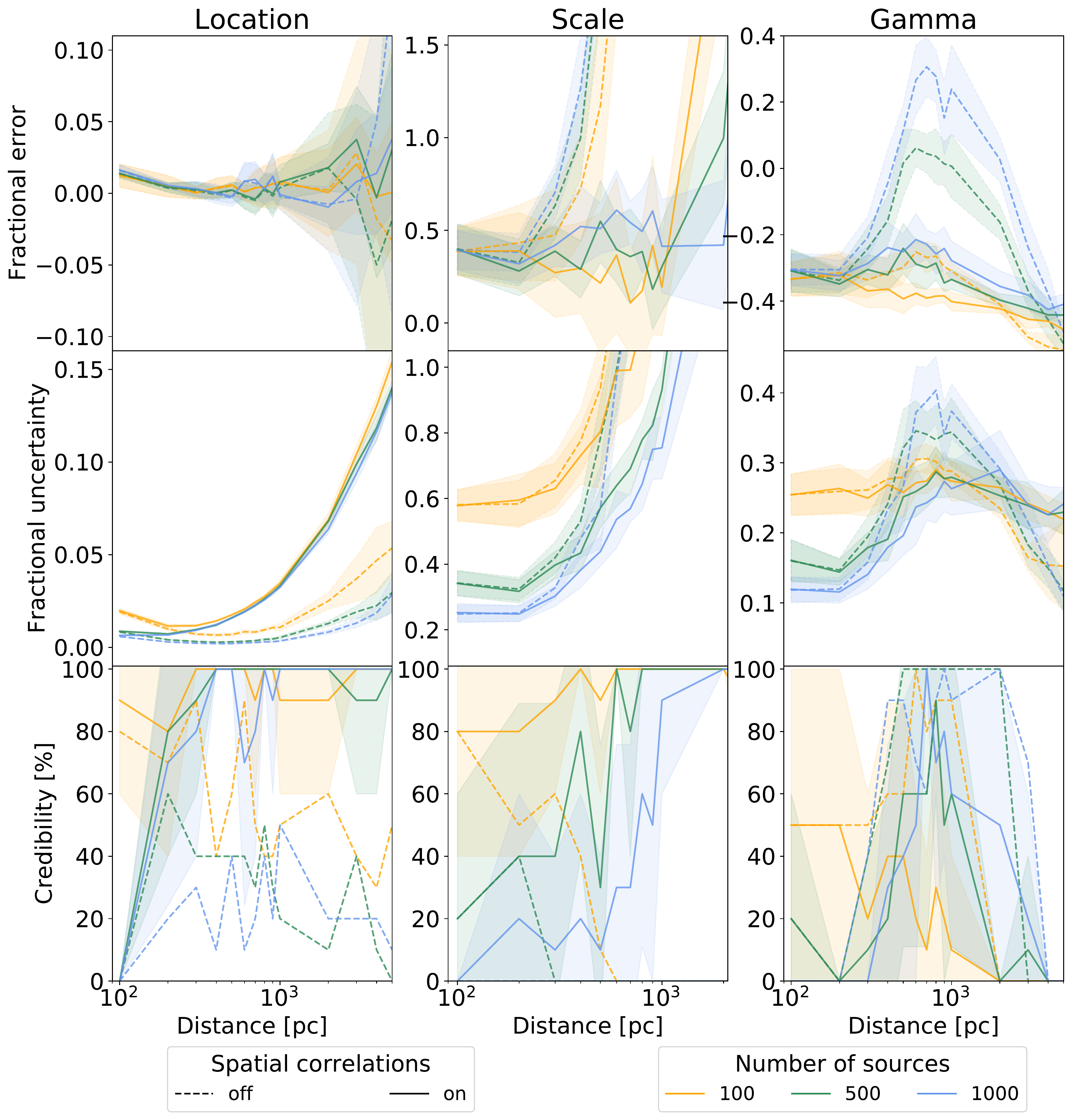}
    \caption{Same as Fig. \ref{fig:A&P_uniform} for the EFF prior.}
    \label{fig:A&P_EFF}
\end{figure}

\begin{figure}[!ht]
    \centering
    \includegraphics[width=\columnwidth,page=1]{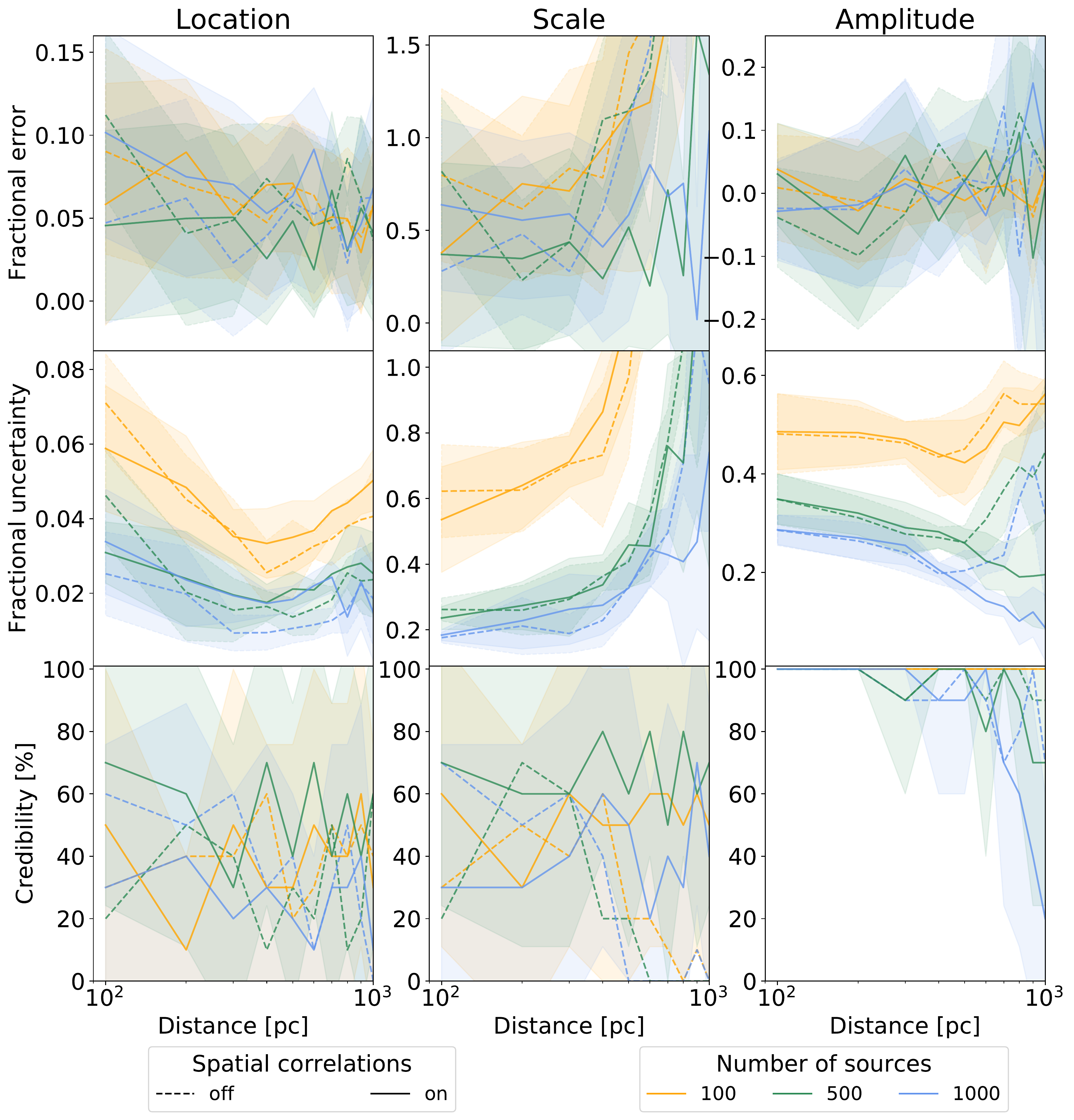}
    \caption{Same as Fig. \ref{fig:A&P_uniform} for the GMM prior. For the sake of clarity, only the parameters of the first component in the mixture are shown. Due to convergence issues, the results of clusters beyond 1 kpc are not shown.}
    \label{fig:A&P_GMM}
\end{figure}

In this section, we present the results obtained after running our methodology using the same prior family that was used to construct the synthetic data set. In our analysis, we use the following configuration. We simultaneously infer the cluster parameters and the source distances using the following hyper-parameters (see Sect. \ref{subsection:hyper_priors}): $\boldsymbol{\alpha}=[\mu,0.1\mu]$ pc, with $\mu$ the distance obtained by inverting the cluster mean parallax, $\beta=100$ pc, $\gamma_{hyp;EFF}=[3,1]$, $\gamma_{hyp;King}=10$, and $\delta=[5,5]$. These hyper-parameters produce weakly informative priors (i.e. the information they provide is smaller than that available) that we expect to cover the diverse cluster information scenarios in which \textit{Kalkayotl} can be used, especially those with weak prior information. In specific cases where more information is available, and thus a more informative prior can be constructed, the performance of our methodology is expected to improve.

We apply the \textit{Kalkayotl} methodology to our synthetic data sets using the minimum number of tuning iterations that ensured convergence. It ranged from 1000 for the Uniform and Gaussian prior, to 10000 for the King prior on the farthest clusters. We use 2000 sampling steps and two parallel chains, which amounted to 4000 samples of the posterior distribution. For our purposes, these are enough samples to compute precise estimates (<2\%) of the posterior statistics. Central and non-central parametrizations were used for the nearby (<500 pc) and far away clusters, respectively. The parallax zero point was set to 0 mas because, as mentioned in Sect. \ref{section:data_sets}, we did not include systematic parallax shifts in the generation of synthetic data. For each synthetic cluster, the inference model was run two times, one includes the parallax spatial correlations and the other neglects them.

\subsection{Population parameters}
\label{subsection:population_parameters}

We now discuss the results obtained at the population level. For each parameter of our cluster prior families, Figures \ref{fig:A&P_uniform} to \ref{fig:A&P_King} show, as a function of cluster distance, the following indicators: i) accuracy in the form of fractional error, which is defined as the posterior mean minus the true value divided by the true value, ii) precision in the form of fractional uncertainty, which is defined as the 95\% credible interval divided by the true value, and iii) credibility, defined as the percentage of synthetic realizations in which the 95\% credible interval includes the true value.

The Uniform prior family recovers the location parameter with excellent accuracy, the fractional error is smaller than 3\%, and its standard deviation < 6\%. The precision is good with a fractional uncertainty smaller than 15\%. The credibility is also good, with more than 80\% of the realizations correctly recovering the true value. Clusters located at less than 300 pc show an increase in the fractional error of the location parameter with respect to those located at 400-500 pc. This decrease in accuracy results from minor violations to our Assumption \ref{ass:proportion}. Concerning the scale parameter, it is accurately recovered, with fractional errors smaller than 10-20\%, only in clusters located closer than 1 kpc, or closer if it has less than 500 sources. The scale precision decreases with distance and improves with the number of sources, as expected. It has large credibility, in the range of 80-90\%, only for clusters located closer than 1 kpc. Beyond this limit, the credibility diminishes as a consequence of the large fractional errors. When the parallax spatial correlations are neglected, we observe the following aspects. First, the accuracy of both parameters decreases. Nonetheless, the accuracy in the cluster distances determination is less affected than that of the scale parameter. Second, the precision of both parameters is underestimated. Third, the credibility of both parameters is severely lowered as a consequence of the larger fractional errors and the underestimated uncertainties.

The Gaussian prior family produces similar results to those of the Uniform one, although with the following differences. First, the precision of the scale parameter remains stable for clusters up to 400 pc, and its absolute value improves with respect to that obtained with the Uniform prior family. Second, when the parallax spatial correlations are neglected the precision of the scale parameter shows better results than in the case of the Uniform prior family. In addition, the credibility of both parameters also increases, in particular for the scale parameter of clusters located closer than 400 pc.

The location and scale parameters of the King prior family behave in a similar way as those of the Gaussian one. Nonetheless, the scale parameter is determined with lower precision. Regarding the tidal radius, it is determined with fractional errors smaller than 10\% only in clusters closer than 500 pc, and with more than 500 sources. Beyond the 500 pc, the results are still credible but noisy, with the errors compensated by the large uncertainties. When the parallax spatial correlations are neglected the results of the location and scale parameters are similar to those of the Gaussian prior family. The tidal radius is underestimated in the region of 300 pc to 2 kpc, and overestimated beyond 2 kpc. The large fractional error and small uncertainty diminish the parameter's credibility down to 20\% at 1 kpc. The increased credibility at 3-4 kpc results from the large uncertainties.

In the EFF prior family, the location parameter is determined with larger noise and uncertainty, with respect to those of the previous prior families. However, it has credibility larger than 80\% for clusters beyond 200 pc. On the contrary, the scale parameter fractional errors are systematically larger, 40\% on average. This large and systematic fractional error reduces the scale credibility even in clusters with 500 and 1000 sources. The Gamma parameter also shows systematic fractional errors but towards smaller values, thus resulting in credibilities smaller than 80\%. Based on these results we conclude that inferring both the scale and gamma parameters simultaneously produces and non-identifiable model. This can be solved by fixing the gamma parameter to obtain a Cauchy or Plummer distributions, in which the accuracy and precision of their parameters improve to values similar to those of the Uniform prior.

In our analysis of the GMM prior family, we use two components, which already make it the most complex of our prior families; with three times more parameters than those of the Uniform and Gaussian prior families. Due to this complexity, we faced difficulties to ensure convergence of the MCMC algorithm in clusters located beyond 1 kpc. Thus, Fig. \ref{fig:A&P_GMM} shows only those cases in which convergence was warranted. In addition, the figure only shows the results of the closest of the two Gaussian components. As can be observed from this figure, both the location and scale parameters are overestimated by 5-10\% and 40-80\%, respectively. This overestimation results from the confusion between the components. Due to the symmetry of this model, its components can be interchanged resulting in locations that are overestimated for the closest component and underestimated for the farthest one. In addition, the scale of both components is overestimated. Despite the issues related to the model symmetry and its lack of identifiability the amplitudes of both components are recovered with low fractional errors and high credibility. The identifiability problem can be partially solved if there is prior information that can be used to break the symmetry\footnote{Further details can be found in \url{https://mc-stan.org/users/documentation/case-studies/identifying_mixture_models.html}}.

\subsection{Source distances}
\label{subsection:source_distances}

\begin{figure}[!ht]
    \centering
    \includegraphics[width=\columnwidth,page=2]{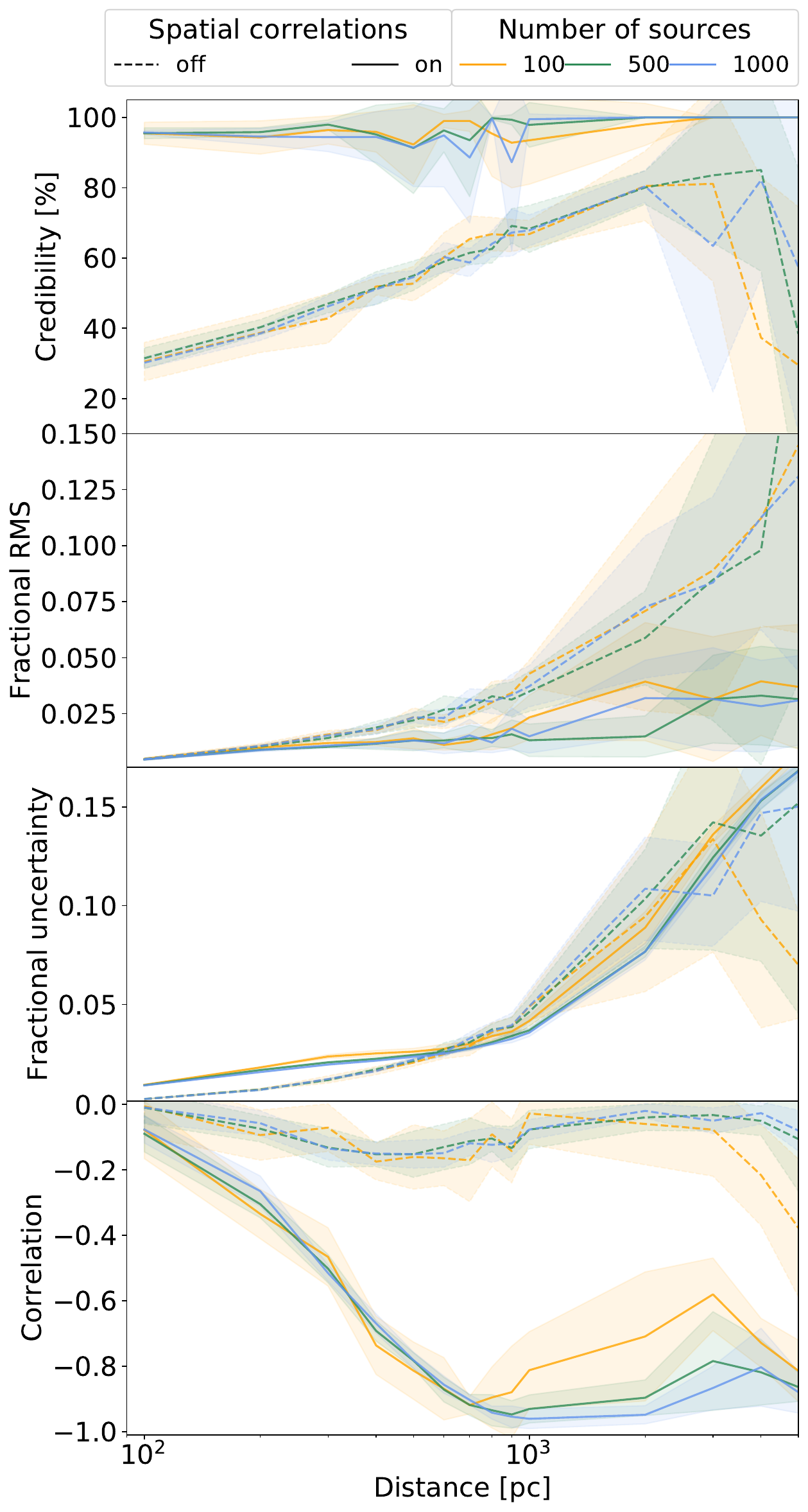}
    \caption{Results of the Uniform prior family. Panels show the credibility, fractional error, fractional uncertainty, and correlation coefficient of source distances as functions of the cluster distance. Captions as in Fig. \ref{fig:A&P_uniform}.}
    \label{fig:source_Uniform}
\end{figure}

We now discuss the performance of our methodology at recovering the individual source distances (i.e. those to the cluster stars). Figure \ref{fig:source_Uniform} shows, at each cluster distance, the mean of the following indicators: i) credibility (as defined above), ii) fractional root-mean-square (hereafter rms) error, iii) fractional uncertainty, and iv) correlation coefficient between the distance error and the offset of the source to the cluster center (more details below). For simplicity reasons we only show the results obtained with the Uniform prior family. The rest of the prior families produce similar results, except for the GMM one in which the credibility diminishes to 60\% for sources in clusters located beyond 700 pc. 

The credibility of our source distance estimates is higher than 90\%, a value that contrasts with that obtained when the parallax spatial correlations are neglected. In this latter case, the credibility increases constantly from 30\% for the closest clusters to a maximum of 80\% for the 3-4 kpc clusters; beyond this latter value, it sinks again. The low credibility obtained when the parallax spatial correlations are neglected is a consequence of the underestimated uncertainties, of both the cluster parameters (see the previous section) and source distances, and the comparatively large fractional errors.

The fractional rms error remains below the 5\% in most of our prior families and almost all cluster distances. The only exceptions are the clusters at 4-5 kpc measured with the EFF and GMM prior families, nonetheless, these have mean fractional rms values lower than 8\%. This indicator also shows the lowest performance when the parallax spatial correlations are neglected. The difference between the fractional rms error obtained with and without the parallax spatial correlations is negligible for the closest clusters (<500 pc) but grows with distance until it reaches 15\% at 5 kpc. 

The mean fractional uncertainty shows two distinct regimes. First, for the clusters closer than 1 kpc it remains low at values < 3\%. Then, it increases with distance and reaches 15\% at 5 kpc. The uncertainties of the individual distances are influenced by the cluster size, in the sense that well defined and compact clusters produce low uncertainties in the stellar distances. Thus, the two observed regimes in the fractional uncertainty are explained as follows. When the cluster scale is accurately estimated, the uncertainties of the individual distances are driven mainly by the parallax uncertainty, which is the case for clusters up to 1 kpc. However, as soon as the scale parameter is overestimated, which occurs beyond 1 kpc, the uncertainties of the individual distances are driven by both the parallax uncertainty and the cluster scale. Since the latter grows with increasing cluster distance, then the uncertainties of the source distances grow as well. Finally, we observe than neglecting the parallax spatial correlations results in uncertainties that are underestimated with respect to the true model for clusters closer than 700 pc, and then overestimated for the rest of the distances. This behavior of the fractional uncertainty results also from the combined influence of the parallax uncertainty itself and the fractional error of the cluster scale. In this case, the fractional error of the latter starts to increase at smaller distances than that observed when the parallax spatial correlations are not neglected (see Fig. \ref{fig:A&P_uniform}).

As discussed in Sect. \ref{subsection:prior_comparison}, the inferred distances to individual sources within a cluster show an error that is anti-correlated with the source position with respect to the cluster center. The value of the anti-correlation coefficient depends on both the parallax uncertainty and the cluster size. Sources with a parallax uncertainty that produces a posterior distance distribution that is narrower than the cluster size have negligible anti-correlation value. Thus, if the precision in the source distance is smaller than the cluster size, then the source position can be accurately determined within the cluster. On the other hand, sources with increasing parallax uncertainties result in posterior distances that are increasingly dominated by the cluster prior, and by the scale parameter in particular. Thus, the mode of the posterior distribution of these sources is attracted to the mode of the prior. Finally, the distances of sources in the near (far) end of the cluster are over(under)-estimated producing thus the anti-correlated error. In all our prior families we observe that the anti-correlation coefficient attains its maximum at 1 kpc, and then it either remains constant for the populous clusters or diminishes for the poorest ones. As described in the previous section, 1 kpc is the limit at which we can accurately estimate the cluster sizes. Therefore, the increase in the anti-correlation coefficient is explained by the continuous increase in the parallax uncertainty at the constant and accurately determined cluster size. Beyond 1 kpc the cluster size is overestimated and the anti-correlation stops growing. Neglecting the parallax spatial correlations results in a lower anti-correlation coefficient. Although it may seem a desirable effect, it is simply explained by an over-estimated cluster size, which for the purposes of this work, is an undesirable effect.

Our main conclusion from this analysis is that, although source distances obtained when the parallax spatial correlations are taken into account have an anti-correlated error, the value of this error is smaller than that of the distances estimated without the parallax spatial correlations. 

\section{Sensitivity to the hyper-parameter values}
\label{appendix:sensitivity}

\begin{figure}[!ht]
    \centering
    \includegraphics[width=\columnwidth,page=1]{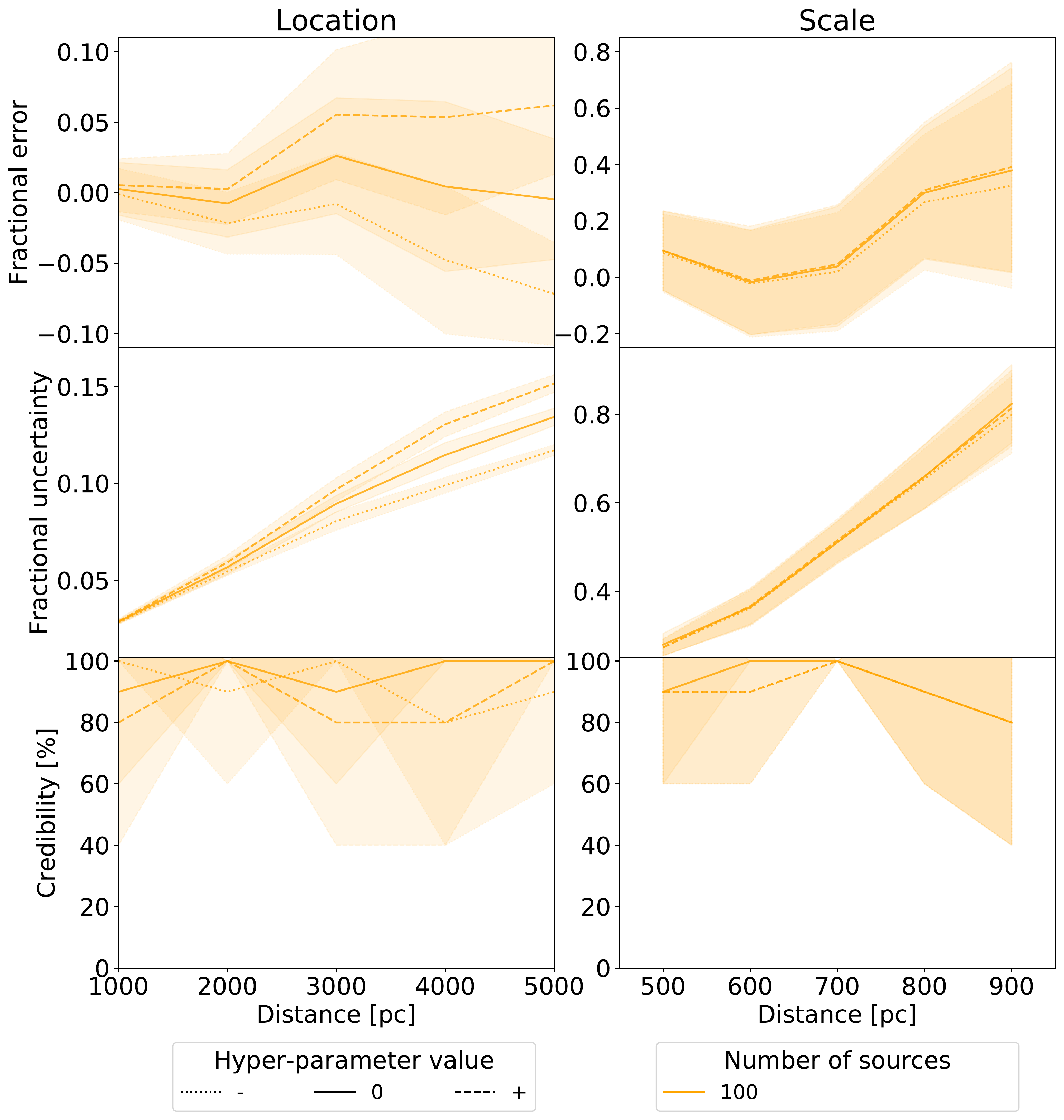}
    \caption{Same as Fig. \ref{fig:A&P_Gaussian} but now the line styles indicate the different hyper-parameter values: + and - for the corresponding $\mu'=\mu(1\pm0.1)$ and $\beta'=\beta(1\pm0.5)$, while 0 for the original values (see text).}
    \label{fig:sensitivity}
\end{figure}

The inference of model parameters is more influenced by the prior, and therefore sensitive to its hyper-parameter values, under poorly constraining data sets. Thus, we reassess the accuracy, precision, and credibility of both the location and scale parameters on the less informative of our data sets: those of clusters with  100 sources and located at the farthest distances: 1 to 5 kpc. In addition, since the scale parameter is only accurately determined at distances closer than 1 kpc, we analyze its sensitivity to the hyper-parameter values in clusters at distances of 500 to 900 pc. For simplicity, we only present the sensitivity analysis performed for the Gaussian prior family.

In Sects. \ref{subsection:population_parameters} and \ref{subsection:source_distances} we use $\boldsymbol{\alpha}=[\mu,0.1\mu]$, with $\mu$ the distance obtained by inverting the cluster mean parallax, and $\beta=100$ pc as hyper-parameters of the location and scale parameters, respectively. To evaluate the sensitivity of our methodology to these hyper-parameters, we change their values to $\boldsymbol{\alpha}'=[\mu',0.1\mu']$ with $\mu'=\mu(1\pm0.1)$ and $\beta'=\beta(1\pm0.5)$. The latter implies evaluating the sensitivity of our methodology to offsets in the hyper-parameter values of 10\% in location and 50\% in scale. In general, hyper-parameters are often set using the information available \textit{a priori}. Thus, we chose the previous offset percentages since: i) we do not expect large variation in the estimates of the cluster distance obtained by simply inverting its mean parallax, and ii) we do expect considerable variations in the estimates of cluster sizes obtained from the literature (see for example Table 1 of \citealt{2018A&A...612A..70O}). 

Figure \ref{fig:sensitivity} shows the fractional error, fractional uncertainty, and credibility of the location and scale parameters as a function of distance. As can be observed, the location parameter is insensitive to the change of its hyper-parameter values up to 4 kpc. Beyond this latter value, the variations due to the hyper-parameter values start to be larger than those due to random fluctuations (i.e. those introduced by the ten randomly simulated data sets of each cluster). Similarly, the variations in the fractional uncertainty due to hyper-parameter values are larger than the random fluctuations only at 4 kpc and beyond. Finally, the credibility of the location parameter is also negligibly affected by the hyper-parameter values since it remains larger than 80\%. The scale parameter is even more insensitive to its hyper-parameter values. In all analyzed distances, the variations in fractional error and uncertainty introduced by changes of 50\% in the hyper-parameter values are all contained within the fluctuations produced by the random sampling of the cluster members. Furthermore, the credibility of this parameter is almost unaffected. 

Our conclusion from this section is that, within the range of analyzed hyper-parameter values, our methodology remains insensitive in clusters located up to 4 kpc. 

\section{Validity of the inverse mean parallax}
\label{appendix:naive}

\begin{figure}[ht]
    \centering
    \includegraphics[width=\columnwidth,page=1]{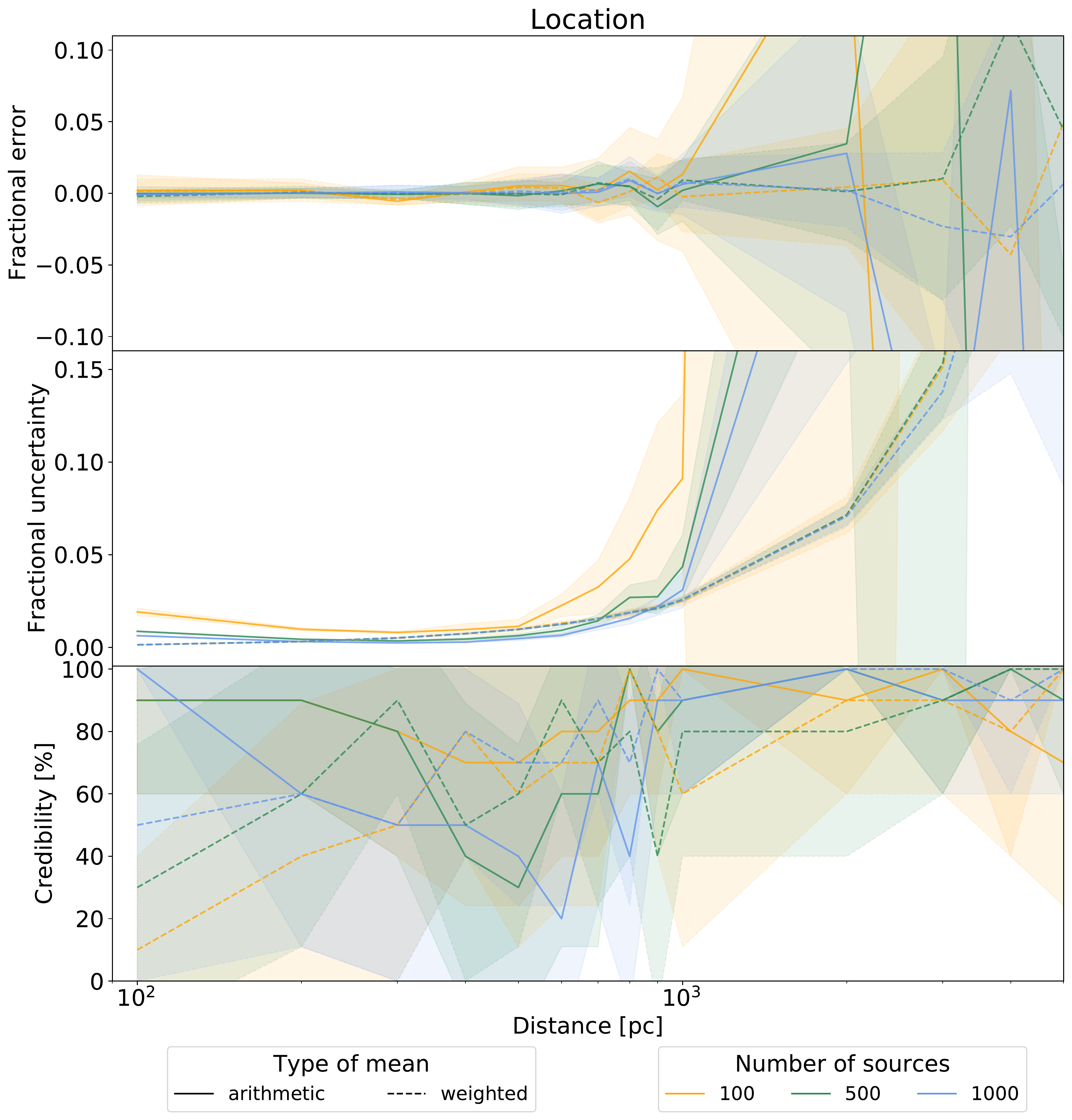}
    \caption{Fractional error, fractional uncertainty, and credibility of the cluster distance estimated by inverting the mean parallax. The lines show two types of mean estimates: arithmetic and weighted (see text). The rest of the captions are as those of Fig.\ref{fig:A&P_Gaussian}. }
    \label{fig:naive}
\end{figure}
In this Appendix, we explore the validity of the common approach in which the distance to a cluster is estimated as the inverse of the mean parallax of its stars. We use the two most common ways of computing the mean of the stellar parallaxes: the arithmetic mean and the weighted mean. While the former does not take into account the observational uncertainties, the latter use them to assigns weights to the individual parallaxes. Here we set these weights as the inverse variance of the observational uncertainties. The uncertainty of these estimators is given by their standard error (i.e. standard error of the mean and standard error of the weighted mean). Thus, we obtain cluster distances as follows. First, we compute the mean (arithmetic or weighted) parallax and its standard error. Second, the cluster distance is computed as the inverse of the mean parallax. Third, the uncertainty of the cluster distance is computed as the 95\% percentile (for compatibility with the results obtained in Appendix \ref{appendix:quality}) of the inverse of one thousand samples drawn from a Gaussian distribution with location and scale as the mean parallax and its standard error, respectively.

We applied the procedure described above to our set of synthetic clusters (see Sect. \ref{section:data_sets}). Fig. \ref{fig:naive} shows the fractional error, fractional uncertainty, and credibility of the cluster distance (i.e. location parameter) for the case of Gaussian distributed synthetic clusters. Results are similar for clusters following other distributions, with the exception of those generated using the GMM prior family, as expected. As can be seen from the figure, both mean estimates return fractional errors that are negligible (<5\%) for clusters located closer than 1 kpc in the case of the arithmetic mean, and 2 kpc in the case of the weighted mean. Beyond these limits, the random fluctuations of the fractional error rapidly reach values larger than 10\%. The fractional uncertainty follows a similar pattern, with values >10\% for clusters farther away than 1 kpc, in the case of the arithmetic mean, and 2 kpc, in the case of the weighted mean. The credibility plot shows the following. The weighted mean, due to its smaller uncertainties, results in low credibility (<50\%) distance estimates to the nearest clusters. Nonetheless, it grows and reaches values greater than 90\% for clusters farther away than 1 kpc. The arithmetic mean results in large credibility ($\sim$80\%) distance estimates for nearest clusters (at 100-200 pc), but then this credibility diminishes and reaches its minimum at $\sim$500-700 pc. Then it grows again (>90\%) for clusters farther away than 1 kpc.

From the previous analysis, we observe the following. First, inverting the mean parallax results in cluster distance estimates with small fractional errors (<5\%) for clusters located closer than 1 or 2 kpc, the latter depends on the type of mean. Second, the uncertainties of the distance estimates obtained with the inverse mean parallax are larger than 10\% for clusters farther away than 1 kpc. Third, the credibility of these distance estimates is lower for nearby clusters than for those farther than 1 kpc. Therefore, we conclude that this approach results in accurate (fractional error <5\%) distances estimates with low credibility (<80\%) for nearby clusters (<1 kpc), and high credibility (>80\%) but low accuracy (fractional errors >10\%) in faraway clusters (> 1 kpc).

\end{appendix}
\end{document}